\pdfoutput=1
\documentclass[12pt]{article}
\usepackage[square,numbers]{natbib}
\usepackage[margin=1.25in]{geometry}

\usepackage[utf8]{inputenc}
\usepackage[T1]{fontenc}
\usepackage{url}            
\usepackage{booktabs}       
\usepackage{amsfonts}       
\usepackage{nicefrac}       
\usepackage{microtype}      
\usepackage{xcolor}

\usepackage{amsmath}
\usepackage{wrapfig}

\usepackage{booktabs}
\usepackage{float}

\usepackage{adjustbox}
\usepackage{tablefootnote}
\usepackage{caption}
\usepackage{bbm}
\usepackage{verbatim}
\usepackage{amsmath}
\usepackage[ruled,vlined]{algorithm2e}

\usepackage{enumitem}
\usepackage{makecell}

\usepackage{xurl}
\usepackage{lineno}
\usepackage{setspace}

\usepackage{todonotes}
\usepackage{framed}
\usepackage{hyperref}
\hypersetup{colorlinks,
            linkcolor=blue,
            citecolor=blue,
            urlcolor=blue,
            linktocpage,
            plainpages=false}
\usepackage[section]{placeins}
\DeclareMathOperator*{\argminA}{arg\,min}

\date{}
\title{\textbf{\Large Behavior measures are predicted by how information \\ is encoded in an individual's brain}} 

\author{
\textbf{Jennifer Williams} \\
Computational Biology Dept.\\
Carnegie Mellon University \\
jlw1@cs.cmu.edu \\
\and
\textbf{Leila Wehbe} \\
Machine Learning Dept. \\
Neuroscience Institute \\
Carnegie Mellon University \\
lwehbe@cs.cmu.edu \\
}
\begin{document}
\maketitle
\begin{abstract}
Similar to how differences in the proficiency of the cardiovascular and musculoskeletal system predict an individual's athletic ability, differences in how the same brain region encodes information across individuals may explain their behavior. However, when studying how the brain encodes information, researchers choose different neuroimaging tasks (e.g., language or motor tasks), which can rely on processing different types of information and can modulate different brain regions. We hypothesize that individual differences in how information is encoded in the brain are task-specific and predict different behavior measures. We propose a framework using encoding-models to identify individual differences in brain encoding and test if these differences can predict behavior. We evaluate our framework using task functional magnetic resonance imaging data. Our results indicate that individual differences revealed by encoding-models are a powerful tool for predicting behavior, and that researchers should optimize their choice of task and encoding-model for their behavior of interest.
\end{abstract}

\section*{Introduction}
Linking individual brain differences to behavior measures can increase our understanding of the brain-behavior relationship. Recent neuroimaging studies have demonstrated that individual differences in neuroimaging measures are predictive of behavior, including measures of brain anatomy\textsuperscript{\citep{Cox2019StructuralBiobank, Kristanto2020PredictingConnections}}, functional connectivity (FC)\textsuperscript{\citep{Kristanto2020PredictingConnections, Finn2015FunctionalConnectivity, Li2019GlobalBehavior, Finn2021Movie-watchingBehavior}}, and structural connectivity (SC)\textsuperscript{\citep{Kristanto2020PredictingConnections, Powell2018LocalFactors, Genc2018DiffusionIntelligence}}. We hypothesize that behavior can also be predicted by how the same brain region, in different individuals, encodes information. As an analogy, athletic ability is not only related to the size of the components of the cardiovascular and musculoskeletal systems or the strength of the connections between components, it is also related to the proficiency of the individual components\textsuperscript{\citep{Maron2006TheAthletes}}. 

Given our hypothesis, we expect that individual differences in how information is encoded in the brain will predict behavior. Additionally, since neuroimaging tasks require the processing of different types of information and can modulate different brain regions, we also expect that individual differences in the encoding of information in the brain will be task-specific, and predict different behaviors. Just like we might ask an athlete to run to measure their endurance, and lift weights to measure their strength, experimenters interested in predicting behavior should tailor their choice of task to the behavior of interest.

An increasingly popular approach to study how information is encoded in the brain is to use encoding-models\textsuperscript{\citep{Naselaris2011EncodingFMRIe, Vu2011EncodingModels, Sudre2012TrackingAccess, Khosla2020AResponse}}. Encoding-models allow researchers to test hypotheses about what information is processed in different brain regions.
Researchers design feature-spaces that approximate stimulus attributes of interest (e.g., visual features of images, syntactic properties of words, etc.). Then, they train an encoding-model to predict the brain activity in each region (e.g., voxel, region of interest (ROI), source), as a function of a specific feature-space. 
The predictive performance of this encoding-model can help identify brain regions that are modulated by the stimulus attributes of interest (those captured by the selected feature-space). Further, in contrast to neuroimaging approaches such as connectivity measures which capture the relationship between brain regions, encoding-models capture a relationship between stimulus attributes and brain activity, which can be generalized to and validated on new stimuli. 

\begin{figure}[t!]
\begin{framed}
    \centering
    \caption{\textbf{Predicting behavior from encoding-model performance}}
    \includegraphics[width=1\textwidth]{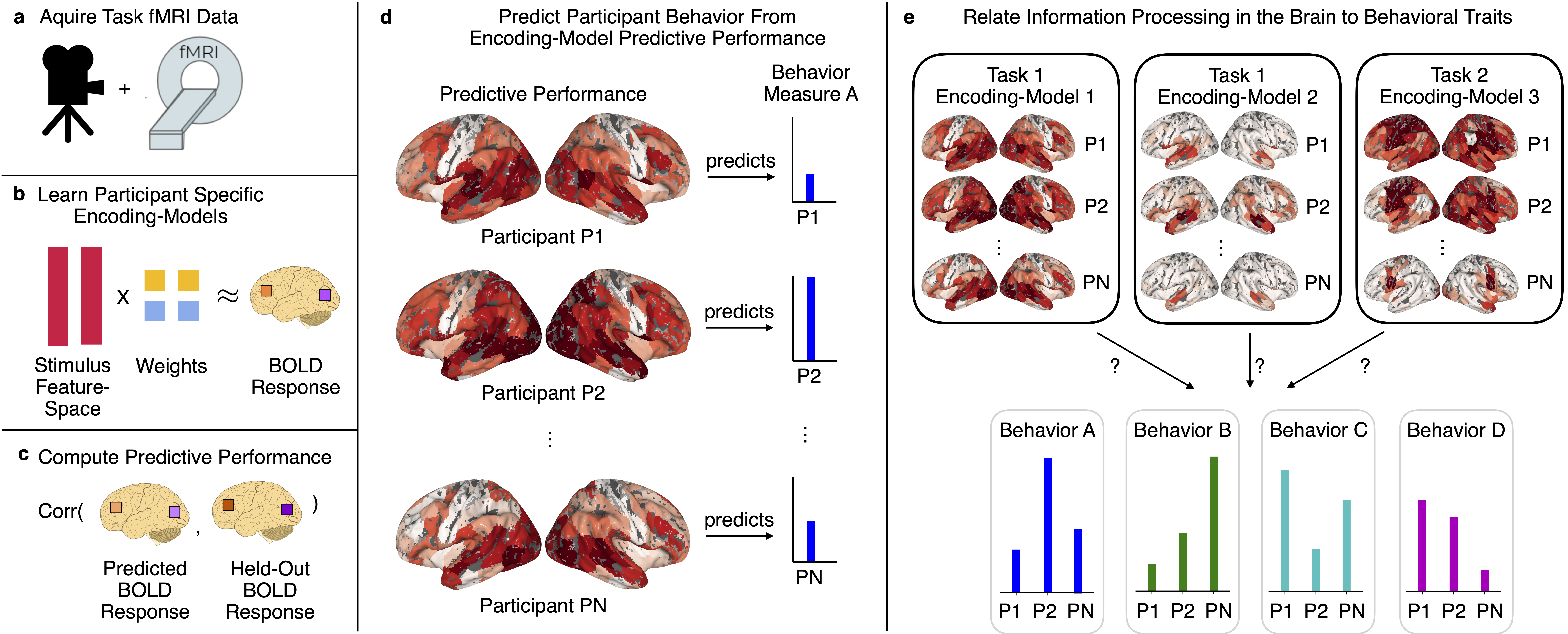}
    \begin{flushleft}
    \small 
    \textbf{a}, The same task functional magnetic resonance imaging (fMRI) data are acquired from all participants. \textbf{b}, Participant-specific encoding-models are learned, which predict the task fMRI blood-oxygen-level-dependent (BOLD) response of each brain region from the stimulus feature-space. \textbf{c}, Predictive performance is estimated as the correlation between the actual and the encoding-models' predicted BOLD response, on a held-out portion of the experiment. The patterns of predictive performance values across the brain approximate the encoding of information in the brain. \textbf{d}, Encoding-model performance can be used to predict participant behavior measures. \textbf{e}, The framework allows testing of what task/encoding-model combinations predict different behavior measures. Different tasks refer to different activities (i.e., watching video clips) participants perform when their BOLD responses are collected. Different encoding-models refer to the different stimulus feature-spaces (i.e., language semantic features) used to predict the BOLD response. Effectively, the framework relates individual differences in the encoding of information in the brain to individual differences in behavior. 
    \end{flushleft}
    \label{fig:framework}
    \end{framed}
\end{figure}

Researchers have used encoding-models to build complex maps of the rich information underlying processes such as vision\textsuperscript{\cite{Khosla2020AResponse,  Kay2008IdentifyingActivity,Nishimoto2011ReconstructingMovies,Huth2012ABrainc}}, language\textsuperscript{\cite{Mitchell2008PredictingNouns, Wehbe2014SimultaneouslySubprocesses,Huth2016NaturalCortex, Deniz2019TheModality, deHeer2017TheProcessing,Fyshe2019TheBrain}}, or auditory processing\textsuperscript{\cite{Khosla2020AResponse, Norman-Haignere2018NeuralCortex, Santoro2014EncodingCortex}}. These maps reveal what type of information is encoded in each region. These maps can also have high spatial resolution because encoding-models can be trained voxel-wise, in contrast with other methods that are computationally infeasible to perform at the whole brain voxel level such as functional connectivity. Because training robust encoding-models requires a lot of data from a large, diverse stimulus set, a trade-off exists between the number of participants and the amount of data acquired from each participant. Consequently, encoding-models are typically used in studies with a lot of data from a few participants, which typically prevents them from leading to insights into individual differences, since a large number of participants is required to robustly identify individual differences. Recently, datasets containing a lot of data and a large number of participants have been released by consortia such as the Human Connectome Project (HCP)\textsuperscript{\citep{VanEssen2013TheOverview}} and the ABCD Study\textsuperscript{\citep{ABCD}}. These datasets enable encoding-models to be used to study individual differences.  

Here, we propose a framework to identify individual differences that is built on encoding-models (Fig.~\ref{fig:framework}). We use this framework to evaluate our hypothesis that individual differences in brain encoding can predict behavior measures. To evaluate our hypothesis, we determine if individual differences predict participant behavior (Fig.~\ref{fig:framework}d), and if the capability to predict particular behaviors is specific to the task (e.g., watching video clips) and encoding-model feature-space (e.g., language semantic features) (Fig.~\ref{fig:framework}e). We use task functional magnetic resonance imaging (fMRI) blood-oxygen-level-dependent (BOLD) responses collected when 176 participants from the HCP\textsuperscript{\citep{VanEssen2013TheOverview}} watched an hour of naturalistic video clips and when they performed a tightly controlled motor task for seven minutes (Fig.~\ref{fig:framework}a). We use encoding-models to predict the BOLD response of each brain region from the stimulus feature-space\textsuperscript{\citep{Naselaris2011EncodingFMRIe, Vu2011EncodingModels, Sudre2012TrackingAccess}} (Fig.~\ref{fig:framework}b). Then, we identify individual differences in the encoding-models' predictive performance (i.e., the correlation between the actual and predicted BOLD response) (Fig.~\ref{fig:framework}c). 
The patterns of predictive performance values across the brain approximate the encoding of information in the brain. We measure how these patterns vary according to what task is chosen, and what encoding-model is paired with the task. We next interrogate the brain-behavior relationship, by using an individual's encoding-model performance to predict 15 cognitive behavior measures in the HCP\textsuperscript{\citep{VanEssen2013TheOverview}} (Fig.~\ref{fig:framework}d). We focus on cognitive behavior measures because these measures are of interest to researchers studying the brain-behavior relationship, and multiple approaches have been proposed using neuroimaging measures to predict them\textsuperscript{\citep{Kristanto2020PredictingConnections, Finn2015FunctionalConnectivity, Li2019GlobalBehavior, Finn2021Movie-watchingBehavior, Powell2018LocalFactors, Genc2018DiffusionIntelligence}}. Then, we evaluate how different task/encoding-model feature-space combinations impact the capability to predict cognitive measures (Fig.~\ref{fig:framework}e).

Together, our results show that individual differences exist in where and what stimulus information is encoded in the brain, and that these differences are predictive of variability in cognitive behavior. Given what is expected in predicting behavior from fMRI, our results argue that encoding-models are a powerful tool for
studying the brain-behavior relationship\textsuperscript{\citep{Shen2017UsingConnectivity, Cui2018TheFeatures, Grady2021InfluenceData}}. Crucially, our results also reveal that the ability to predict different behavior measures depends on the choice of task and encoding-model.

\section*{Results}
\subsubsection*{Average participant encoding-model (APE) estimation and evaluation} 
We first sought to identify individual differences in brain encoding. We used encoding-models, as they (1) predict the BOLD response of a brain region (i.e., voxel, ROI, source) as a function of the stimulus feature-space and (2) can be computed on an individual participant level. Throughout the paper, we performed our analysis on two levels. We estimated encoding-models for 268 ROIs defined by the Shen functional brain atlas\textsuperscript{\citep{Shen2013GroupwiseIdentification}} for each participant; furthermore, we estimated encoding-models at the voxel level. We constructed an \textbf{average participant encoding-model (APE)} which predicts a participant's BOLD response in every brain region (ROI or voxel) from the BOLD response of all brain regions (ROIs or voxels, respectively) averaged across all other participants. As the stimulus drives the brain activity that is shared across participants\textsuperscript{\citep{Hasson2004IntersubjectVision}}, the average BOLD response is a function of the entire stimulus. Critically, APEs are flexible enough to learn how every brain region in a participant is related to the activity in all brain regions (averaged over all other participants), and thereby account for participant spatial and/or functional individuality. APEs allow us to test hypotheses about the individual differences in BOLD response in every brain region, related to the entire stimulus. 

\begin{figure}[h]
 \begin{framed}
    \centering
    \caption{\textbf{A consistent amount of information is shared across participants, and is very reliable in some areas}} 
    \includegraphics[width=\textwidth]{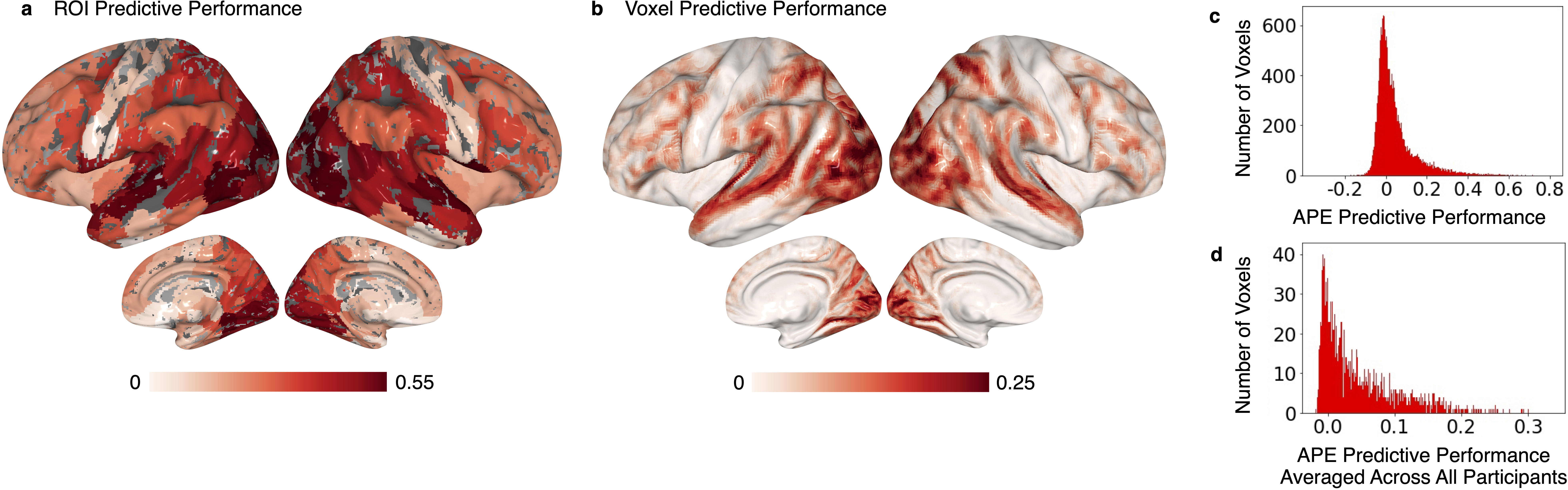}
    \begin{flushleft} 
    \small \textbf{a}-\textbf{b}, Predictive performance of the video clip task average participant encoding-model (APE) averaged across all participants at a (\textbf{a}) region of interest (ROI) and (\textbf{b}) voxel level. Prediction performance of a participant's BOLD response predicted from the average BOLD response of all other participants was obtained on held-out data using a cross-validation setup. Predictive performance is also known as Pearson's r. Prediction performance is high in ROIs in the visual cortex, temporal cortex, as well as parts of the prefrontal and association cortices. Voxel level results align with ROI results but reveal a concentration of voxels with high predictive performance in the early visual cortex and the superior temporal gyrus. \textbf{c}-\textbf{d}, Histograms of the voxel level predictive performance of the video clip task APE (\textbf{c}) for each participant and (\textbf{d}) averaged across all participants. On average, APEs are good predictors of participant level brain activity, indicating that the way that stimulus-driven information is processed is reliably shared across participants.
    \end{flushleft}
    \vspace{-0.15in}
    \label{fig:Consistent}
 \end{framed}
\end{figure}

Before we identified individual differences revealed by APEs, we first tested if APEs are good predictors of a participant's BOLD response. Using ridge regression, we estimated APEs that predict the BOLD response of one participant from the average BOLD response across all other participants. Each time-point of the BOLD response is predicted from the corresponding time-point in the average response. Next, we tested how well the video clip APEs predicted the BOLD response elicited by a new video clip that was not used for model estimation. We visualize the predictive performance (the correlation between the actual and the predicted BOLD response time-series) after averaging it across all participants, at either the ROI or voxel level in Fig.~\ref{fig:Consistent}a,b. We found good predictive performance at the ROI level in the visual, temporal, and parts of the prefrontal and association cortices (Fig.~\ref{fig:Consistent}a). Voxel level results align with the ROI level results but also reveal a concentrated area of high predictive performance in the early visual cortex and the superior temporal gyrus (Fig.~\ref{fig:Consistent}b). We also observe that the predictive performances are lower for the voxel level than the ROI level. A possible explanation for this difference, is that because each ROI's BOLD response is an average of the BOLD response across voxels', the ROI response is less noisy and consequently easier to predict than the voxel response. However, the pattern of high predictive performance at the voxel level is more spatially specific. Given that there is high level of spatial consistency between ROI level and voxel level results, and the fact that the voxel level maintains more spatial specificity, we give priority here and in the rest of the paper to the voxel level results. Thus, we show the distribution of the voxel level predictive performance for each participant (Fig.~\ref{fig:Consistent}c), and averaged over all participants (Fig.~\ref{fig:Consistent}d). We also show for each ROI and voxel, the number of participants for which the predictive performance is significantly higher than chance (one-sided permutation test, alpha = $0.05$, false discovery rate (FDR) corrected\textsuperscript{\citep{Benjamini1995ControllingTesting}}) (Supplementary Fig.~\hyperref[fig:sup_sig_voxel_EM]{1}a,d). Together these results highlight that APEs are generally good predictors of a participant's BOLD response, this indicates that processing of stimulus-driven information is reliably shared across participants. 

\subsubsection*{The variability of encoding-model performance is feature-space specific}
\begin{figure}[t!]
 \begin{framed}
    \centering
    \caption{\textbf{The variability of encoding-model performance is feature-space specific.}}
    \includegraphics[width=1\textwidth]{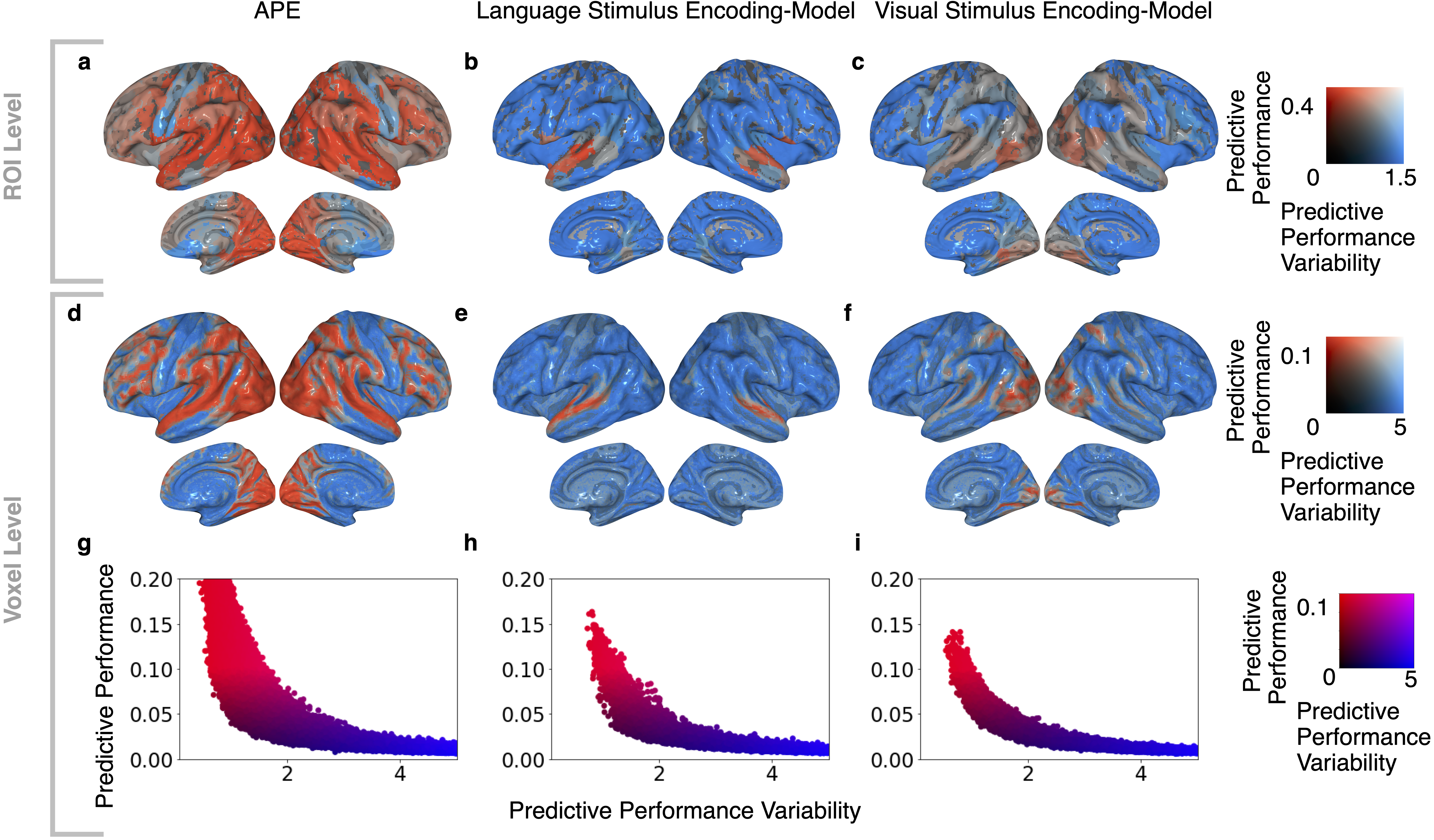}
    \begin{flushleft}
    \small \textbf{a}-\textbf{c}, Comparison of the average predictive performance to its variability for the (\textbf{a,d}) video clip APE, (\textbf{b,e}) language stimulus encoding-model, and (\textbf{c,f}) visual stimulus encoding-model at the ROI (\textbf{a-c}) and voxel level (\textbf{d-f}). The ROIs and voxels are colored according to the top two 2D color maps, where red corresponds to high predictive performance, blue to high variability, white to high predictive performance and variability (no ROIs or voxels are assigned this color), and absence of color to low predictive performance and variability. We find high predictive performance and low variability at both levels for the following: (\textbf{a,d}) APE in previously defined language regions\textsuperscript{\citep{Fedorenko2010NewSubjects}} including the anterior (ATL) and posterior temporal lobes (PTL), the visual cortex, and the dorsal ("where") pathway and the ventral ("what") pathway; (\textbf{b,e}) language stimulus encoding-model in the same previously defined language regions; (\textbf{c,f}) visual stimulus encoding-model in the visual cortex, and the dorsal and ventral pathways, consistent with previous results\textsuperscript{\citep{Huth2012ABrainc}}. \textbf{g}-\textbf{i}, Voxel level scatter plots of variability versus predictive performance. Red voxels correspond to high predictive performance, blue to high variability, purple to high predictive performance and variability (no voxels are assigned this color), and black to low predictive performance and variability (according to the bottom 2D color map). Our results show a strong trend for voxels with high predictive performance to have low variability (red), and voxels with low predictive performance to have high variability (blue).
    \end{flushleft}
    \label{fig:StimFeatSpecific}
 \end{framed}
\end{figure}

We next investigated how encoding-model performance can vary according to the feature-space that is used. We compared the APE predictive performance with that of two other encoding-models with different stimulus feature-spaces. First, we created a language stimulus encoding-model, that predicts a participant's BOLD response from a word embedding space, built from each spoken word in the video clips word2vec embedding, consisting of 300 features that approximate the semantic properties of the word (i.e., words with similar meanings have similar features)\textsuperscript{\citep{word2vec}}. Then, we created a visual stimulus encoding-model, that predicts a participant's BOLD response from a visual semantic space built from the visual features in each frame in the video clips, consisting of 859 binary features that denote which important objects and action categories were present. The visual features were provided in the HCP and created following the method in Huth et al.\textsuperscript{\citep{Huth2012ABrainc}}. We estimated the predictive performance for an encoding-model for each participant, and computed its average across participants. To evaluate if the choice of encoding-model can reveal distinct patterns of individual differences, we also measured the across-participant predictive performance variability (Fig.~\ref{fig:StimFeatSpecific} and Supplementary Fig.~\hyperref[fig:sup_roi_satter]{2}) by calculating the coefficient of variation. For each ROI or voxel, the coefficient of variation is the standard deviation of the predictive performance (across all participants) normalized by the absolute value of the average predictive performance (across all participants).

Next, we show the average and variability of the predictive performance of the video clip APE, language stimulus encoding-model, and visual stimulus encoding-model, at both the ROI (Fig.~\ref{fig:StimFeatSpecific}a-c and Supplementary Fig.~\hyperref[fig:sup_roi_satter]{2}) and voxel level (Fig.~\ref{fig:StimFeatSpecific}d-i). Average predictive performance on its own can be seen in Fig.~\ref{fig:Consistent} and Supplementary Fig.~\hyperref[fig:sup_pred_perf_stimlus_models]{3}. There are no ROIs or voxels with both high average predictive performance and high variability.
ROIs and voxels with high average predictive performance and low variability are shown in red. Using APE, these ROIs and voxels with high average prediction performance and low variability span the visual, temporal, and parts of the prefrontal cortices. Using the language stimulus encoding-models, these ROIs and voxels are constrained to previously defined language ROIs in the anterior and posterior superior temporal lobes (ATL and PTL)\textsuperscript{\citep{Fedorenko2010NewSubjects}}, and using the visual stimulus encoding-model, they are constrained to the early visual cortex, and the dorsal and ventral pathways. The locations of previously defined language ROIs and visual cortex ROIs are shown on the cortical surface in Supplementary Fig.~\hyperref[fig:sup_roi_map]{4} to help the reader with interpretation. For each ROI and voxel, the number of participants for which the predictive performance was significantly higher than chance (one-sided permutation test, alpha = $0.05$, FDR corrected) for the three encoding-models is shown in Supplementary Fig.~\hyperref[fig:sup_sig_voxel_EM]{1}. We conclude that the APE predicts the BOLD response in many regions including the visual cortex and the part of temporal cortex associated with the language system\textsuperscript{\citep{Fedorenko2010NewSubjects, Fedorenko2014ReworkingNetwork}} while the stimulus encoding-models predict a subset of the BOLD response either in the visual or temporal cortex.
 
Our results allow us to make two conclusions. First, these findings support our assumption that APE captures a super-set of the variance of activity across ROIs or voxels captured by the individual stimulus encoding-models. This suggests that brain activity averaged across other participants is a "more comprehensive" representation of the stimulus than current state of the art feature-spaces extracted from the stimulus, at least for predicting BOLD activity.  

Second, we observe that the patterns of variability are not consistent across the three encoding-models. Some ROIs and voxels have high variability and low predictive performance (colored in blue) in one encoding-model but not the other. For example, the same region (e.g., the PTL) might have high variability in a less complete model (e.g., visual stimulus encoding-model), and low variability in a more complete model (APE) or a model that approximates the relevant stimulus attributes (e.g., language stimulus encoding-model). These results highlight that the variability of encoding-model performance depends on the type and amount of information present in each encoding-model. As we are interested in predicting behavior, we hypothesize that the choice of encoding-model should be made while considering the behavior of interest.

\begin{figure}[h!]
 \begin{framed}
    \centering
    \caption{\textbf{The variability of APE predictive performance is task-specific.}}
    \includegraphics[width=0.6\textwidth]{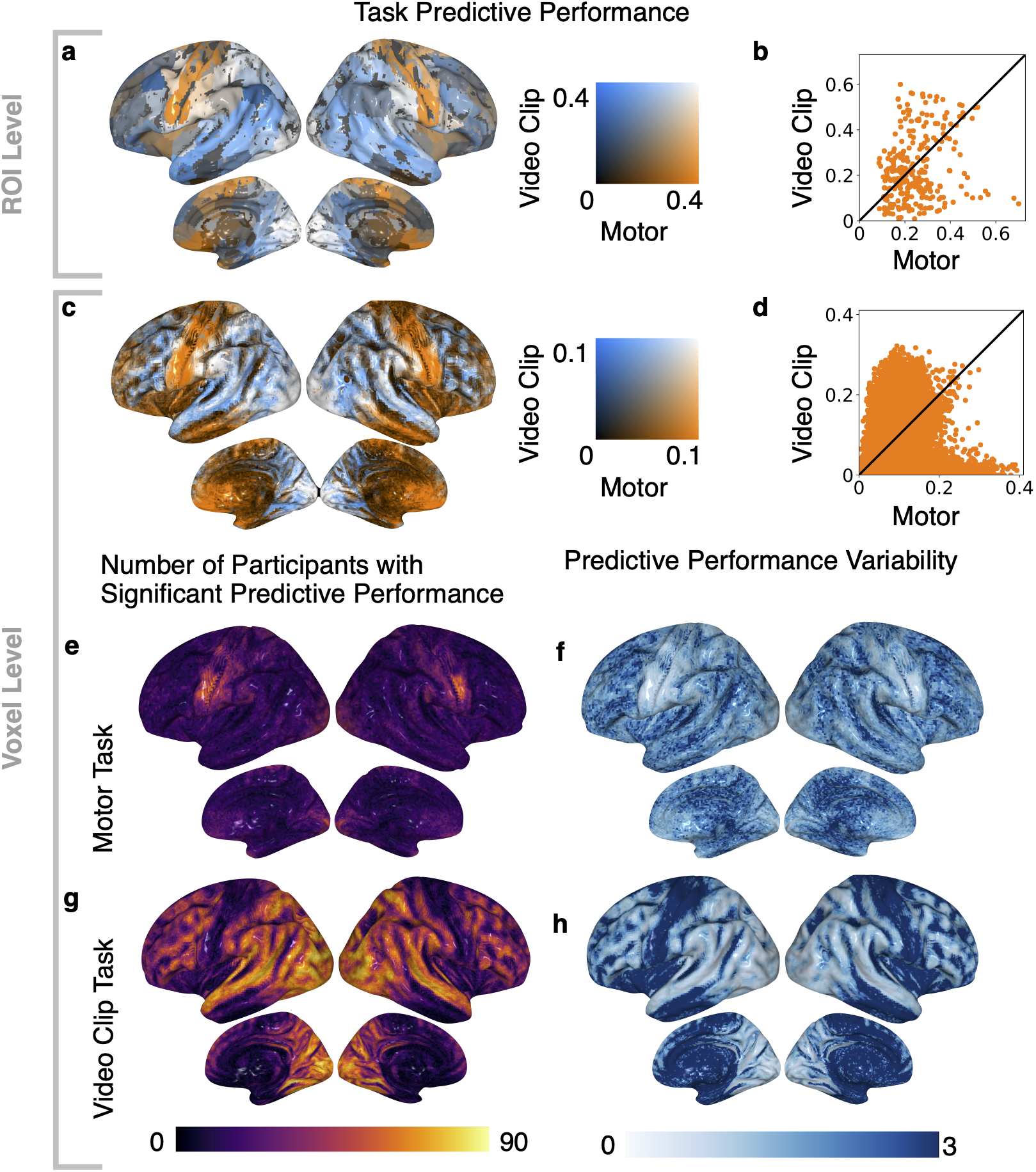}
    \begin{flushleft}
    \small \textbf{a,c}, APE predictive performance for the motor task compared to the video clip task averaged across all participants at the ROI and voxel level on the cortical surface. ROIs and voxels are colored according to the 2D color maps. Good predictive performance is shown at both levels for the following: motor task in the motor cortex (orange); video clip task in the auditory cortex and association areas (blue); both tasks in the visual cortex (white). \textbf{b,d}, In the scatter plots, the same data are plotted per ROI or voxel. The black diagonal line denotes $y=x$; points above this line represent ROIs or voxels whose video clip task predictive performance is greater than their motor task predictive performance. We find that some ROIs/voxels have similar performance for both tasks, while a large portion have high performance for the video clip task but not the motor task, and many have high performance for motor task but not the video clip task. \textbf{e,g}, For each voxel, the number of participants for which the predictive performance is significantly higher than chance (one-sided permutation test, alpha = $0.05$, FDR corrected) is task-specific (motor task (\textbf{e}), video clip task (\textbf{g})). \textbf{f,h}, Per voxel predictive performance variability during the motor task (\textbf{f}) is low in the motor cortex and high elsewhere, but during the video clip task (\textbf{h}), it is low in the visual, temporal, and prefrontal cortices and high elsewhere. Predictive performance magnitude and variability are task-specific. 
    \end{flushleft}
    \vspace{-0.15in}
    \label{fig:StimSpecific}
 \end{framed}
\end{figure}
    
\subsubsection*{The variability of APE predictive performance is task-specific} 
We noticed that the motor cortex has high variability of predictive performance from the video clip APE (Fig.~\ref{fig:StimFeatSpecific}a,d and Supplementary Fig.~\hyperref[fig:sup_voxel_feature_specific_non_norm_var]{5}a,d). Moreover, we observed a low number of participants whose performance in the motor cortex was significantly higher than chance (Fig.~\ref{fig:StimSpecific}g and Supplementary Fig.~\hyperref[fig:sup_sig_voxel_EM]{1}a,d). Given both observations, we hypothesized that the high variability in the motor cortex was not a special property of the motor cortex but that it was due to the choice of task. Specifically, our hypothesis was that the task did not activate the motor cortex enough to generate reliable activity. To test this, we created a motor task APE, which predicts a participant's BOLD response associated with performing a motor task (from HCP) from the average activity of all other participants. We compared the average predictive performance of the motor task APE and the video clip task APE at the ROI and voxel level on the cortical surface in Fig.~\ref{fig:StimSpecific}a,c, and in scatter plots in Fig.~\ref{fig:StimSpecific}b,d. We found that some ROIs/voxels have similar predictive performance for both tasks, while a large portion have high performance for the video clip task but not the motor task, and many have high performance for motor task but not the video clip task (Fig.~\ref{fig:StimSpecific}b,d). We observed good average predictive performance for the motor task in the motor cortex (orange); the video clip task in the auditory cortex and association areas (blue); and both the motor and video clip tasks in the visual cortex (white) (Fig.~\ref{fig:StimSpecific}a,c). The good predictive performance of the visual cortex for the motor task was expected as the participants were presented with visual cues during the motor task. Fig.~\ref{fig:StimSpecific}e,g, shows, for each voxel, the number of participants for which the predictive performance is significantly higher than chance (one-sided permutation test, alpha = $0.05$, FDR corrected) for both the motor and video clip tasks (ROI level results in Supplementary Fig.~\hyperref[fig:sup_roi_motor]{6}a,c) are consistent with these results). The task-specific performance magnitude suggests that the use of APEs with different tasks reveal task-specific individual differences. 

To further show that the high variability in the motor cortex during video clip task was due to the choice of task, we compare the predictive performance variability between the video clip and motor tasks. We measured the variability across participants and found that across ROIs and voxels, the variability of the motor task APE is not correlated with that of the video clip task APE (ROI level: correlation = $0.04$, \textit{p} = $0.54$, voxel level: correlation = $-0.0004$, \textit{p} = $0.93$) (Supplementary Fig.~\hyperref[fig:supp_motor_movie_var]{7}). We also found that, for both tasks, ROIs and voxels with high predictive performance have low variability, and ROIs and voxels with low predictive performance have high variability (Fig.~\ref{fig:StimSpecific}a,c,f,h, Supplementary Fig.~\hyperref[fig:sup_roi_motor]{6}b,d). Further, the motor cortex has high predictive performance and low variability for the motor task and low predictive performance and high variability for the video clip task. 
The task effect on the measured variability is important because it suggests that a task that does not activate an area reliably can lead to very different patterns of variability than a task that does. Because the estimated variability is different given the task, we hypothesize that, in addition to the choice of encoding-model, the choice of task should be made while considering the behavior of interest. For instance, one hypothesis is that, to predict motor proficiency, a task that activates the relevant region (i.e., motor cortex) and reveals individual differences in the region's function might do better than a task that does not activate this region. Conversely, another hypothesis is that a task that doesn't usually recruit the motor cortex (such as the video clip task) might reveal individual differences in the few people for which the motor cortex is recruited by this task (e.g., people with high ability to imagine), which might allow us to predict certain cognitive measures. 

\subsubsection*{Encoding-model performance is related to behavior}
Encoding-models allow us to estimate predictive performance at the individual level. However, it is still unclear if individual differences in predictive performance are related to behavior, and if the choice of task and encoding-model impact this relationship. We used 15 cognitive measures to study how a range of cognitive processes are related to individual differences revealed by encoding-models. We present the correlation of the encoding-model performance with scores from 15 cognitive measures for each voxel in the four encoding-models in Fig.~\ref{fig:behavior} (four measures) and in Supplementary Fig.~\hyperref[fig:sup_corr_cogn_voxel]{8}

\begin{figure}[h]
 \begin{framed}
    \centering
    \caption{\textbf{Cognitive behavior is associated with individual differences in encoding-model performance.}}
    \includegraphics[width=1\textwidth]{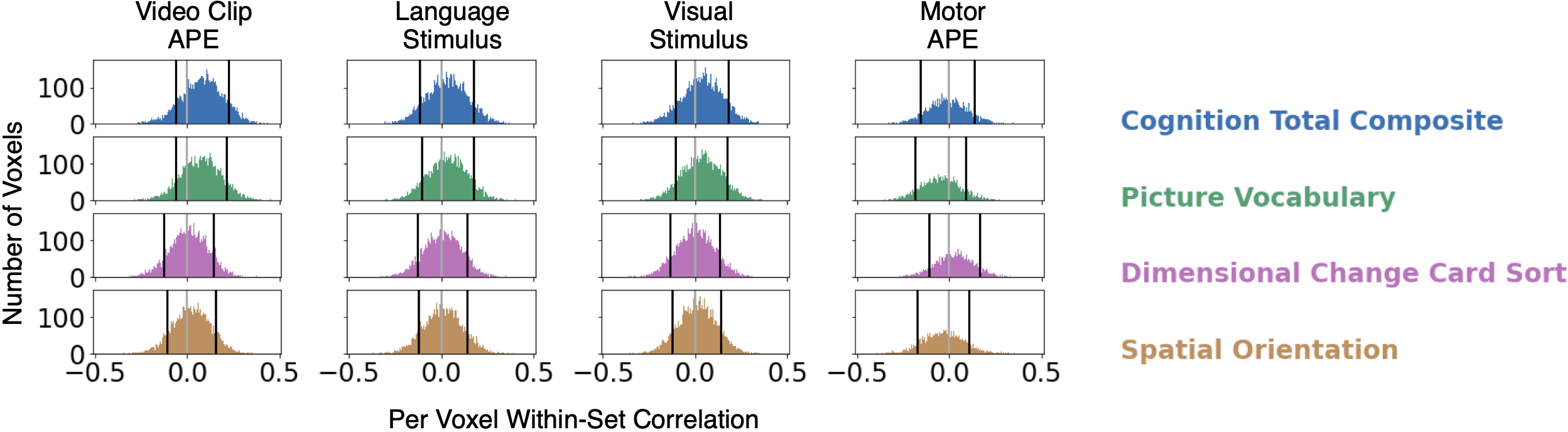}
    \begin{flushleft}
    \small Correlation of encoding-model performance with scores from four cognitive measures. We use all four encoding-models (three for the video clip task and one for the motor task). For each voxel, the predictive performance across participants is correlated with their cognitive scores. Histograms of the per-voxel correlation measures are shown, with black lines denoting the 10th and 90th percentiles. The light gray line denotes zero correlation. We observe variability in the distributions of correlations across different encoding-model/cognitive measure combinations. For some encoding-model/cognitive measure combinations, there is a clear positive skew of the correlations. For example, there is a positive correlation in most of the brain between a participant's video clip APE predictive performance and their \emph{cognition total composite} score. However, the pattern between the same encoding-model's performance and the participants' \emph{spatial orientation} scores is inconsistent.
    \end{flushleft}
    \label{fig:behavior}
\end{framed}
\end{figure}

\noindent
(remaining 11 measures). Results at the ROI level are presented in Supplementary Fig.~\hyperref[fig:sup_corr_cogn_roi]{9}. Detailed descriptions of the cognitive measures are provided in Supplementary Table \hyperref[tab:sup_cogn]{1}. These measures included both composite and individual measures of cognitive processes including executive function, memory, attention, language, and spatial orientation. We found that all of the cognitive measures are correlated (or anti-correlated) with the encoding-model performance for a least some voxels within each of the four encoding-models (at least some voxels have an absolute value of correlation higher than $0.4$ in each case). We also found variability in the distributions of the correlations across different encoding-model/cognitive measure combinations. For example, there was a positive correlation in most of the brain between a participant's video clip APE predictive performance and their \emph{cognitive total composite} score, but the pattern between the same encoding-model's performance and the participants' \emph{spatial orientation} scores was inconsistent. Together this suggests that encoding-model performance may provide new insights into the relationship between individual differences in the representation of information in the brain and behavior, and that the behavior of interest may need to be considered when choosing the encoding-model and task.

We subsequently examined if individual differences in encoding-model performance predict individual differences in cognitive behavior, when evaluated in a more stringent, out-of-set setup, where the model is evaluated on held-out data. We created behavior models which predict a participant's cognitive measure score from the pattern of predictive performance over ROIs or voxels for a specific encoding-model. We used ridge regression to estimate these behavior models. Additionally, we tested how well the behavior models predict participant cognitive measure scores on held-out participants not used for the model estimation, using leave-one-family-out-cross-validation. These models allowed us to test hypotheses about how individual differences in brain encoding are related to individual differences in behavior. 

\begin{figure}[h!]
 \begin{framed}
    \centering
    \caption{\textbf{Individual differences in encoding-model performance predict cognitive behavior.}} 
    \includegraphics[width=0.8\textwidth]{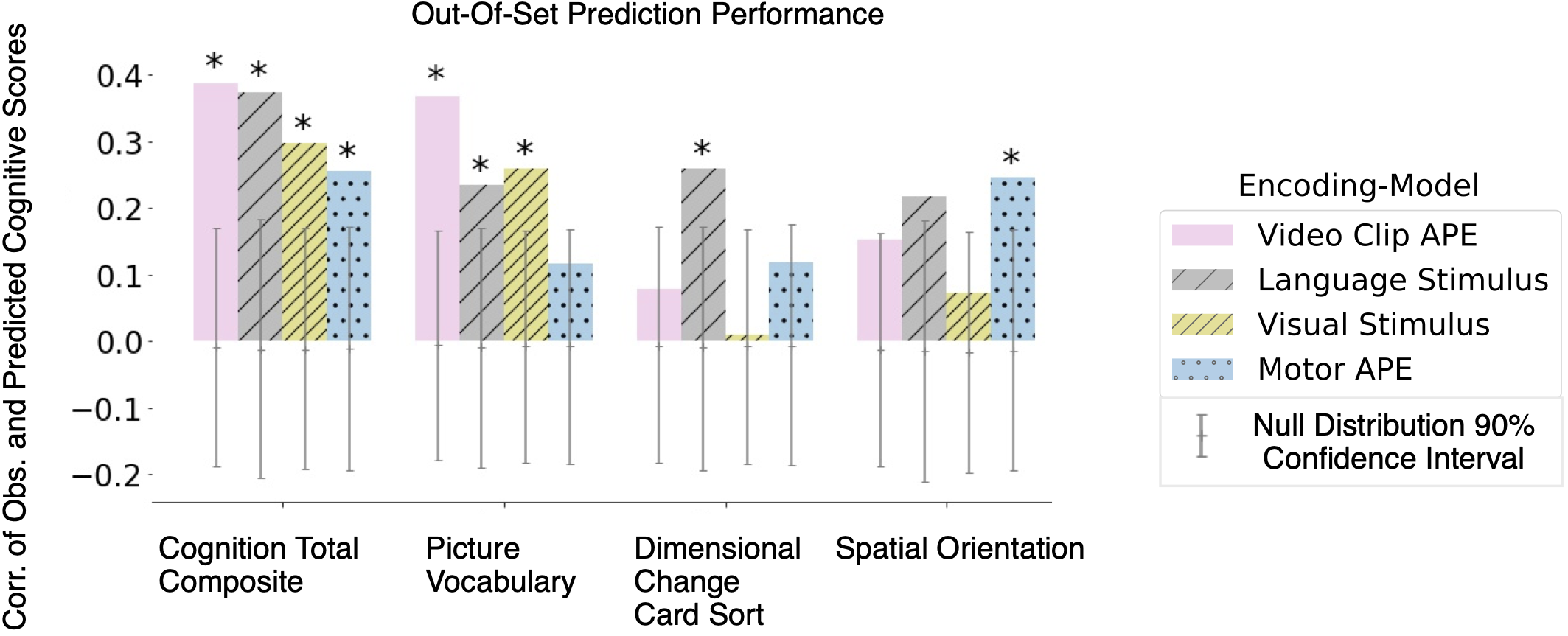}
    \begin{flushleft}
    \small Out-of-set predictive performance of participant cognitive scores. Performance was measured as the correlation between actual and predicted participant cognitive measure scores. Each participant's score was predicted from a vector of voxel encoding-model performance (using all four encoding-models). Predictive performance was evaluated on held-out participants in a leave-one-family-out-cross-validation setup. We find that all four encoding-models' performances predict the \emph{cognition total composite} measure significantly higher than chance. We also find that the three-video clip task encoding-models’ performances predict the \emph{picture vocabulary} measure significantly higher than chance. The language stimulus encoding-models' performance predicts the \emph{dimensional change card sort} measure significantly higher than chance. The motor task encoding-models’ performance predicts the \emph{spatial orientation} measure (variable short penn line orientation) significantly higher than chance. The ability to predict behavior measures thus appears to be encoding-model specific and task specific.
    \end{flushleft}
    \label{fig:behavior_pred}
\end{framed}
\end{figure}

Fig.~\ref{fig:behavior_pred} and Supplementary Fig.~\hyperref[fig:behav_predict_voxel_11]{10} present each behavior model's performance, the

\noindent
correlation between the actual and the predicted behavior measure scores at the voxel level (corresponding plots for the ROI level are in Supplementary Figs.~\hyperref[fig:behavior_pred_roi_4_measures]{11} and \hyperref[fig:behav_predict_roi_11]{12}). We found that, at the voxel level, we can predict the scores for most (9/15) of the cognitive measures from at least one of the four types of encoding-models significantly higher than chance (one-sided permutation test, alpha = $0.05$, FDR corrected). This suggests that individual differences in how information is encoded in the brain, as approximated by encoding-model performance, can predict individual differences in a range of cognitive processes. Results for the ROI level were very similar to the voxel level results (correlation = $0.72$, \textit{p} = $10^{-10}$), however the magnitude was not found to be significantly higher than chance after multiple comparison correction, likely due to a loss of power when averaging the voxel data into ROIs. 

To further examine the behavioral relevance of individual differences in brain encoding, we next investigated if different task/encoding-model combinations, at a voxel level, could predict different behavior measures. We found that all four encoding-models’ performances predict the \emph{cognition total composite} measure significantly higher than chance (Fig.~\ref{fig:behavior_pred}). We also found differences in which measures are predicted significantly higher than chance for the three video-clip task encoding-models' and the motor task encoding-model. For example, we found that some measures including the \emph{picture vocabulary} measure are predicted by the three-video clip task encoding-models significantly higher than chance (Fig.~\ref{fig:behavior_pred}, Supplementary Fig.~\hyperref[fig:behav_predict_voxel_11]{10}). Meanwhile, we found that the \emph{spatial orientation} measure (variable short penn line orientation) is significantly predicted by the motor task encoding-model (Fig. \ref{fig:behavior_pred}). We further find differences in which measures are predicted significantly higher than chance by the three-video clip task encoding-models. For example, the language stimulus encoding-models' performance predicts the \emph{dimensional change card sort} measure significantly higher than chance (Fig.~\ref{fig:behavior_pred}). The ability to predict behavior measures thus appears to be encoding-model specific and task specific. This supports our hypothesis that both the task and encoding-model should be chosen while considering the behavior of interest. 

The analyses presented in the results were performed on a development set of 90 participants from the HCP (see \hyperref[methods:participants]{Methods} for details on selection). For reproducibility purposes, we reserved the remaining 86 participants for the final results, after all our analyses have been fixed. Currently, we await reviewer feedback to update and finalize our framework before we extend our analysis to all participants (Total 176; 106 female, age 22-36) in the HCP with 7T fMRI video clip task and resting state data.  

\section*{Discussion}
Our findings support our hypothesis that individual differences in brain encoding are task-specific and predict different behavior measures. We used different task/feature-space combinations to create encoding-models and identified individual differences in their predictive performance. Then we used encoding-model performance as an approximation of the encoding of information in the brain. We demonstrated that individual differences in encoding-model performance are task and feature-space specific. Furthermore, we showed that predicting variability in cognitive behavior from individual differences in encoding-model performance is possible, and that this predictive capability is specific to the choice of task and encoding-model feature-space. These results highlight (1) a new framework to interrogate the relationship between differences in brain encoding and behavior, and (2) that researchers should chose their task and encoding-model while considering their behavior of interest.

We note a few additional considerations about our framework. Firstly, as encoding-models predict BOLD response as a function of stimulus feature-spaces, these models, and consequently our framework, are limited by how well the feature-spaces represent a specific type of information about the stimulus. However, the ability to construct stimulus feature-spaces is constantly improving due to progress in machine learning and artificial intelligence, which will strengthen our framework. Secondly, our framework requires the data from each participant to be in a shared anatomical space, with the option of using ROIs defined across participants, a common procedure in neuroimaging analysis. In this work, we built on prior individual differences work and applied our framework to data in MNI space and used the 268 ROIs defined by the Shen atlas\textsuperscript{\citep{Finn2015FunctionalConnectivity, Shen2013GroupwiseIdentification}}. We observed that voxel level analysis, while having lower encoding-model performance, led to better ability to predict behavior than performing analysis at the ROI level. One of the reasons for this limitation at the ROI level is that the choice of atlas-defined brain regions may suffer from topological variability (i.e., variability in functional–anatomical correspondence) between participants\textsuperscript{\citep{Salehi2020ThereTask, Yaakub2020OnDatabases, Bohland2009TheParcellations}}. Further, it might be that even though our voxel level analysis has higher performance than the ROI level analysis, that it also suffers from the same limitation  because it requires that every participant's data be normalized to MNI space. A future extension of our work can address this limitation by not using a standardized space or atlas-defined ROIs and instead relying on approaches such as hyperalignment\textsuperscript{\cite{haxby2011common}}, regularized canonical correlation analysis\textsuperscript{\cite{bilenko2016pyrcca}} or shared response models\textsuperscript{\cite{ chen2015reduced}}, that can be adapted to estimate (linearly or non-linearly) a shared space from participants with different native spaces. Indeed, hyperalignment has been shown to resolve participant topographic idiosyncrasies while recovering reliable fine-grained (e.g., voxel level) individual differences (in FC and other functional indices)\textsuperscript{\cite{Feilong2018ReliableArchitecture}}. Since our observations are relatively consistent between ROI and voxel level results, we do not expect that addressing this limitation will drastically change the crux of our findings, but it might lead to improved ability to predict behavior.

\subsubsection*{Video clip and motor task APEs differ in ability to predict motor cortex activity} 
One striking finding is that, in the motor cortex, the variability of prediction performance across participants is high in the video clip task, but not in the motor task,
showing that the variability of performance in a given brain area is stimulus specific. Further, we observed high average performance in the motor cortex in the motor task, and not the video clip task. This is in line with prior work that shows that executing a motor action activates a consistent BOLD response in the motor cortex, which might allow the encoding-model to learn how these actions map to the BOLD response in the brain\textsuperscript{\citep{Buccino2001ActionStudy, Gazzola2009TheData}}. However, prior studies have also shown that both executing and observing motor actions activate a consistent BOLD response in the motor cortex, yet we found that watching naturalistic video clips (which includes observing actions) does not activate a consistent response in the motor cortex\textsuperscript{\citep{Buccino2001ActionStudy, Gazzola2009TheData}}. Our results add support to the growing body of work that - sometimes incidentally - shows that observing actions in naturalistic video clips does not consistently activate the motor cortex\textsuperscript{\citep{Huth2016NaturalCortex, Hasson2010ReliabilityStimulation, Moraczewski2018Inter-subjectChildhood, Nguyen2017DistinctViewing}}. Even though there are specific instances where observing segments of naturalistic video clips can consistently activate the motor cortex, such as when a participant observes tennis serves or intricate hand movements\textsuperscript{\citep{Hasson2004IntersubjectVision, Wright2007BrainStudy}}. Importantly, most previous work studying observing motor actions used tightly controlled video clips rather than naturalistic video clips, making it difficult to discern the impact of observing a motor action as part of a complex and cluttered visual environment on the BOLD response in the motor cortex\textsuperscript{\citep{Buccino2001ActionStudy, Rizzolatti2001NeurophysiologicalAction, Shmuelof2006ACortex, Moriuchi2017PrimaryProperties, Li2020BrainActions}}. A possible explanation for support on both sides of the debate is that observing motor actions that are the focus of a video clip, might lead to sustained activation of the motor cortex, which leads to a BOLD response with a high signal-to-noise ratio. Conversely, observing motor actions which are only a minor part of a video clip might lead to a signal-to-noise ratio that is not high enough to consistently reveal motor cortex activation. This emphasizes that the choice of task impacts what information is being reliably processed, and consequently that the choice of task affects what hypotheses can be tested with sufficient statistical power. 

\subsubsection*{Differences in brain encoding predict variability in cognitive behavior}
Notably, individual differences in brain encoding are good at predicting variability in cognitive behavior, given what is expected for predicting behavior from neuroimaging measures\textsuperscript{\citep{Shen2017UsingConnectivity, Cui2018TheFeatures, Grady2021InfluenceData}}. It is difficult to compare different approaches that predict behavior from neuroimaging measures, because approaches use data from a variety of fMRI paradigms (rest, a naturalistic video clip task, and tightly controlled tasks), different numbers of participants (20-1,000+), and analyze the data with different pre-processing choices and prediction methods. However some attempts have been made to determine how well individual neuroimaging measures can predict behavior. For instance some work has shown that the maximum predictive performance for different behaviors stabilizes between 0.2-0.5 for 20-700 HCP participants\textsuperscript{\citep{ Cui2018TheFeatures, Grady2021InfluenceData}}. In addition to our prediction performances for 90 participants being within this range, our performances are as high as the performance range for other neuroimaging measures from many more participants (>700) and single run data\textsuperscript{\citep{Powell2018LocalFactors,Cui2018TheFeatures}}. This suggests that the encoding of information in the brain is at least as informative as other neuroimaging measures for understanding cognition. 

We further found that the predictive capability of individual differences in brain encoding is task and cognition measure specific. Recent work has shown incidental contradictory findings about whether FC's ability to predict behavior is task-specific\textsuperscript{\citep{Tomasi2020NetworkAdults, Jiang2020Task-inducedRelationships}}. As these were incidental findings, it is impossible to make a direct comparison due to both the limited details with respect to these findings and the vast methodological differences between the works. Despite this, these works highlight the growing interest in using task fMRI data to predict behavior and the importance of understanding if the ability to predict behavior from neuroimaging measures is task-specific. Our results strongly suggest that the choice of task should be a crucial consideration. 

\subsubsection*{Predicting behavior yields insights into brain-behavior relationship}
Our encoding-model framework uniquely allows us to learn the link between behavior and individual differences in brain activity related to the entire stimulus (i.e., video clips) while ignoring other aspects of brain activity (noise and non-stimulus activity) using an APE, or parts of the stimulus defined by different feature-spaces (i.e., language and visual) using stimulus encoding-models. Our framework also allows us to learn the link between behavior and individual differences in the brain activity of each brain region. Therefore, our approach captures individual differences that are inherently different from those in anatomy, FC, SC, and ISC. (See Methods sections \hyperref[ISC_methods]{ISC} and \hyperref[FCSC_methods]{Definitions of FC and SC}) for definitions. The distinct individual differences captured by encoding-models and the other neuroimaging measures raise the question, do these measures capture complementary information about the brain? An interesting future direction would be to test if these neuroimaging measures are complementary in their ability to predict behavior and provide insights into the brain-behavior relationship. 

\subsubsection*{Conclusion}
Our findings open the door for neuroimaging studies to increase our understanding of the brain-behavior relationship by investigating the relationship between individual differences in brain encoding and behavior. Further, the task/encoding-model specificity suggests that experimenters interested in predicting behavior should tailor their choice of task and encoding-model to the behavior of interest.

\newpage
\section*{Data availability}
We used publicly available human neuroimaging datasets from the HCP\textsuperscript{\citep{VanEssen2013TheOverview}}. The data can be downloaded by anyone who agrees to the Open Access Data Use Terms (\url{https://db.humanconnectome.org}). Some of the HCP demographic and behavioral data requires acceptance of the HCP Restricted Data Use Terms (\url{https://www.humanconnectome.org/study/hcp-young-adult/document/restricted-data-usage}). To analyze the neuroimaging data at an ROI level, we used the 268 Shen atlas\textsuperscript{\citep{Finn2015FunctionalConnectivity, Shen2013GroupwiseIdentification}}, this is publicly available on the BioImage Suite NITRC page (\url{https://www.nitrc.org/frs/?group\_id=51}). To label Shen atlas ROIs as visual ROIs we used the publicly available labels provided on the BioImage Suite Web (\url{bioimagesuiteweb.github.io/webapp/connviewer.html})\textsuperscript{\citep{Shen2017UsingConnectivity}}. To label the Shen atlas ROIs as language ROIs we used the publicly available previously identified language-relevant regions by Fedorenko et al.\textsuperscript{\citep{Fedorenko2010NewSubjects}}, which were provided from \url{https://evlab.mit.edu/funcloc/download-parcels}.

\section*{Code availability}
The python and bash scripts used to analyze and visualize the neuroimaging data as part of this study are openly available at \url{https://github.com/brainML/great-apes}. 

\newpage
\section*{Methods}
\subsubsection*{HCP data} 
\paragraph{Participants:}
\label{methods:participants}
We analyze publicly available data from the HCP S1200 release\textsuperscript{\citep{VanEssen2013TheOverview}}, with healthy participants between 22-37 years old. The S1200 release contains 184 participants with 7T fMRI data; we restrict our analysis to 176 participants (106 female, age 22-36) for whom all four runs of the video clip task and resting state 7T fMRI data were available. To enable a final test of the reproducibility of the results we split the participants into a development set and a final test set (not used in this version of the paper and reserved for after reviewer comments are accounted for). As most participants have at least one twin or sibling in the data set, to control for the relatedness we ensured that all members of each of the 90 families were in the same set. We further control for gender and age differences, by balancing the gender and then age between the development set (n= 90, 54 female, age 22-36) and test set (n= 86, 52 female, age 22-35). 

\paragraph{fMRI data:}
We use fMRI data from three different paradigms in this study. 
\begin{enumerate}
    \item Four 7T task fMRI (tfMRI) runs (HCP files: tfMRI\_MOVIE1, tfMRI\_MOVIE2, tfMRI\_MOVIE3, tfMRI\_MOVIE4) were acquired while participants watch naturalistic video clips. The video clips include segments from a variety of movies including Star Wars Episode V: The Empire Strikes Back and Ocean's Eleven.  Each tfMRI run is just over 15 minutes long, for a total of 60 minutes and 55 seconds of recorded tfMRI data.
    \item Four 7T resting state fMRI (rsfMRI) runs (rfMRI\_REST1, rfMRI\_REST2, rfMRI\_REST3, rfMRI\_REST4), were acquired while participants fixated at a bright cross-hair on a dark background. Each rsfMRI run is 15 minutes long, for a total of 60 minutes of recorded rsfMRI data. 
    \item Two 3T tfMRI runs (tfMRI\_MOTOR\_LR , tfMRI\_MOTOR\_RL), were acquired while participants performed a tightly controlled motor task. Each tfMRI run is three minutes and 34 seconds long, for a total of seven minutes and eight seconds of recorded tfMRI data. 
\end{enumerate}

Both the 7T rsfMRI and tfMRI scans were acquired with a repetition time (TR) of 1000 ms, echo time (TE) of 22.2 ms, and voxel size of 1.6 mm isotropic. The 3T tfMRI scans were acquired with TR of 720 ms, TE of 33.1 ms, and voxel size of 2 mm isotropic. Additional data acquisition details are provided in Glasser et al.\textsuperscript{\citep{Glasser2013TheProject}}. 

We use minimally preprocessed fMRI from the HCP data repository\textsuperscript{\citep{Glasser2013TheProject}}. The preprocessing steps include motion correction, spatial artifact/distortion elimination and registration to MNI ICBM 152 nonlinear sixth generation space. The 7T fMRI data in addition to being minimally preprocessed is denoised with FSL's FMRIB's ICA-based X-noiseifier (FIX)\textsuperscript{\citep{Griffanti2014ICA-basedImaging, Salimi-Khorshidi2014AutomaticClassifiers}}. Additional details on data processing are provided in \citep{Glasser2013TheProject, Griffanti2014ICA-basedImaging, Salimi-Khorshidi2014AutomaticClassifiers}. 

Two of the participants in the development set did not have both runs of the motor task, and therefore were excluded from the motor task analyses. 

\paragraph{Behavior data:}
We use 15 cognitive measures in this study. These measures assess a variety of cognitive processes including executive function, memory, attention and language. Detailed descriptions of the cognitive measures are provided in Supplementary Table \hyperref[tab:sup_cogn]{1}, and additional details are provided in Barch et al.\textsuperscript{\citep{Barch2013FunctionBehavior.}}. These measures include one measure from each of the 12 cognitive tests conducted in the HCP, and three out of four of the composite measures of cognition. The fourth composite measure, NIH Toolbox \emph{cognition early childhood composite}, was excluded as it is recommended for ages three to six. We used the age-adjusted scale scores for all 10 of the NIH toolbox measures, as this normalizes participant scores based on individuals within the same age group and we want to predict cognition across a normative range.

A few participants in the development set did not have all 15 cognitive measures, and therefore those participants were excluded from the corresponding behavior analyses. Specifically, four participants were excluded from the \emph{cognition fluid} and \emph{cognition total composite} analyses, and one participant was excluded from the \emph{cognition crystallized composite} and \emph{dimensional change card sort} analyses.

\subsubsection*{Preparing fMRI data for ROI and voxel level analysis} 
\paragraph{ROI parcellation:} 
We use the 268 functionally defined ROI Shen atlas\textsuperscript{\citep{Finn2015FunctionalConnectivity, Shen2013GroupwiseIdentification}} to parcellate the brain. As the Shen atlas was previously constructed on a separate dataset of healthy participants, these ROIs were defined independent of our data. The Shen atlas parcellation image is publicly available on the BioImage Suite NITRC page.

First, we resample the Shen atlas (MNI127) to the same template space as our fMRI data (MNI152). Then we downsample each fMRI run for each participant by averaging the voxel BOLD responses within the 268 functionally defined ROIs from the Shen atlas per time frame, similarly to previous work\textsuperscript{\citep{Rosenberg2015AConnectivity, Greene2018Task-inducedTraits, Gao2019CombiningMeasures, Doss2020TheBrain}}. For each fMRI run and each participant this generates a new dataset of size number of TRs by 268 ROIs.  

\paragraph{Voxel selection:}
We use the Pycortex software\textsuperscript{\citep{Gao2015Pycortex:FMRI}} to select the voxels in our 3T (67,427 voxels) and 7T (131,906 voxels) fMRI data for our voxel level analysis. We select the voxels within 2mm of the fiducial surface, to only retain those with high probability of being gray matter, using the "thin" mask option in Pycortex. The segmentation of the MNI template into gray and white matter is done using the Freesurfer software.

\subsubsection*{Average participant encoding-model (APE) matrix}
Our APEs use as input the average activity from all other participants. For each paradigm (video clip task, motor task and resting state) we create an average participant matrix that contains the BOLD response of all brain regions for that paradigm averaged across all other participants. The matrix has as dimensions the number of TRs by the number of brain regions. While this matrix would not typically be considered a feature-space matrix, for the purpose of our modeling it can be thought of as a feature-space as this matrix captures TR specific information about the stimulus.

\paragraph{Resting-state APE:}
To demonstrate that (1) our APE models are strictly stimulus-based and not applicable to resting-state paradigms and (2) the identified individual differences are not an effect of pre-processing or scanner noise we created an APE for the HCP resting-state task. These results are shown in Supplementary Fig.~\hyperref[fig:sup_rest]{13}.

\paragraph{APE compared to ISC:}
To evaluate the difference between APEs accounting for participant spatial and/or functional differences, and ISC assuming a similar organization for different participants' brains, we compare the two types of models. We compare the APE predictive performance and ISC for the video clip task. These results are shown in Supplementary Fig.~\hyperref[fig:sup_ISC]{14}.

\subsubsection*{Language and visual semantic feature matrix}
We create these feature-space matrices to be used as input features for the language stimulus and visual stimulus encoding-models.
 
\paragraph{Language semantic features:}
For each video clip we construct a language semantic feature matrix. We first create time-stamped transcripts that indicates the onset of each spoken word in each video clip in a semi-automated fashion. We obtain written transcripts (not stamped for word onset) of each video clip in the following fashion. For 6/15 video clips we extracted subtitle files from their respective full length movies. We then converted these subtitle files to SRT subtitle files, using gotranscript's subtitle converter. Subtitles were not available for the remaining video clips, therefore we created waveform (WAV) audio files from the video (MP4) files provided in the HCP using iMovie. Then we downsampled the audio files to 1 channel, 16 bits and 16 Khz, using sox. Next we use Google Clouds Speech-to-Text API to obtain SRT subtitle files of the video clips from the audio files.         

We next used Google Cloud's Speech-to-Text API to obtain time-stamped transcripts that indicate the word onset of each video clip, similarly to the approach used by the Courtois Project on neuronal modeling (\url{https://www.cneuromod.ca/}). For each video clip we use the API to generate the time-stamped transcript from the video clip's audio file and transcript (not stamped for word onset). These transcripts were visually inspected and errors were corrected using the Subtitle Edit software. 

Then for each video clip for each word in our transcript we obtain the 300 feature word2vec embedding\textsuperscript{\citep{word2vec}} for that word. We use the pre-trained version of word2vec trained on part of the Google News corpus (approximately 100 billion words) provided by Mikolov et al.\textsuperscript{\citep{word2vec}}. Next, we use a Lanczos filter with the same parameters as Huth et al.\textsuperscript{\citep{Huth2016NaturalCortex}} to downsample each video clips' word embeddings into a language semantic feature matrix where each row corresponds to a 300 feature vector for a TR.

\paragraph{Visual semantic features:} 
For each video clip we create a visual semantic feature matrix. We use the visual semantic features that were provided in the HCP. These features were manually annotated following the method in Huth et al.\textsuperscript{\citep{Huth2012ABrainc}}, where objects presented in each frame were identified and then the objects were labeled with a 859-dimensional binary vector corresponding to their super-categories in WordNet. We create a visual semantic feature matrix where each row is the union of the vectors of the objects occurring in that TR.

\subsubsection*{Encoding-models}
We estimate participant specific encoding-models which predict the brain activity of a brain region (i.e., ROI, voxel) as a function of a feature-space matrix. We use ridge regression (L2-regularized regression) to estimate the encoding-models, similar to previous work\textsuperscript{\citep{Sudre2012TrackingAccess, Nishimoto2011ReconstructingMovies, Wehbe2014SimultaneouslySubprocesses, Huth2016NaturalCortex, wehbe-etal-2014-aligning, Toneva2019InterpretingBrain}}. For each brain region the ridge regularization parameter is selected independently using nested 10-fold cross-validation. Ridge regression is computationally efficient, and it has previously been shown that different regularization approaches lead to similar results for fMRI data if proper regularization is used and the regularization parameter is selected independently using cross-validation\textsuperscript{\citep{Wehbe2015RegularizedSmoothing}}. 

For each brain region $i$, we fit encoding-models to predict the vector of brain activity ${\bf y}^{i} = [y_{1}^{i}, y_{2}^{i}, y_{3}^{i}...y_{N}^{i}]^\top$, where $N$ is the number of TRs, as a function of a feature-space matrix ${\bf X}$. (In this notation, we follow the convention of writing a vector as a column, hence the transpose operations. Vectors are lowercase and in bold font, matrices are uppercase and in bold font). Each encoding-model $m$ is associated with a feature-space in which each time point $t$ is associated with a vector ${\bf x}_{t}^{m}$. This vector contains a fixed number of features $F_m$ that describe aspects of the stimulus at time $t$. The feature-space matrix can thus be written as ${\bf X}^{m} = [{\bf x}_{1}^{m}, {\bf x}_{2}^{m}, {\bf x}_{3}^{m}\ldots{\bf x}_{N}^{m}]^\top$, which is aligned to the brain activity, and where each row corresponds to a TR. To account for the lag in the hemodynamic response in fMRI data, we follow a common approach in building encoding-models \textsuperscript{\citep{Nishimoto2011ReconstructingMovies,Wehbe2014SimultaneouslySubprocesses, Huth2016NaturalCortex, Toneva2019InterpretingBrain}}. We use as input to the encoding-model the feature-space ${\bf Z}^m$ in which the row corresponding to TR $t$ contains the concatenated features of the last nine seconds: ${\bf z}_{t}^m = [{\bf x}_{t}^{m\top},{\bf x}_{t-1}^{m\top}, {\bf x}_{t-2}^{m\top} \ldots {\bf x}_{t-8}^{m\top}]^\top$. For the APE model, since both the input and the output are fMRI activity, we do not delay the matrices. Instead, for each participant $p$, we use ${\bf Z}^m = {\bf \bar Y}_{-p}$, where ${\bf \bar Y}_{-p}$ indicates the average brain activity of all participants except for participant $p$.

For each participant $p$ and each feature-space matrix ${\bf Z}^{m}$, we estimate how well the feature-space predicts the BOLD response in each brain region $i$ using 4-fold-cross-validation. The four folds for the video clip task and resting state paradigms are the four fMRI runs. The four folds for the motor task paradigm are each repetition of the task, this is possible as the task was performed twice within each fMRI run. For each fold: 
\begin{enumerate}
    \item We split the feature-space matrix ${\bf Z}^{m}$ and fMRI data ${\bf Y}_p$ into training and validation matrices. For each matrix, we normalize (mean = 0, standard deviation = 1) each brain region (or each feature) across time. We obtain training matrices ${\bf Z}^{R,m}$, ${\bf Y}_p^{R}$ and validation matrices ${\bf Z}^{V,m}$, ${\bf Y}_p^{V}$.
    \item We use the training matrices to estimate a model ${\bf W}_p^{m}$:
   $$ \argminA_{{\bf W}_p^{m}} || {\bf Y}^R_p - {\bf Z}^{R,m}{\bf W}_p^{m} ||_{2}^{2} + \boldsymbol\lambda||{\bf W}_p^{m}||_{2}^{2} $$
    Crucially, we first identify the regularization parameters, $\lambda^{i}$  that minimizes the nested 10-fold cross-validation error for each brain region $i$. The $\lambda^{i}$ are thus chosen \emph{independently} for each region, allowing regions that are easy to predict to have less regularized weights, and shrinking the weights in regions that are difficult to predict\textsuperscript{\citep{Wehbe2015RegularizedSmoothing}}. After the $\lambda^{i}$s are chosen, 
    we  estimate $\hat {\bf W}_p^{m}$ using all the training matrices. 
    \item We predict the validation fMRI data as: $\hat{\bf Y}_p^{V,m} = {\bf Z}^{V,m}\hat{\bf W}_p^{m}$. 
\end{enumerate}

We then compute the prediction performance for each input matrix, participant, brain region combination, by correlating the concatenation of the predicted fMRI activity across all 4-folds with the observed fMRI activity across the same folds (i.e., Pearson correlation is performed across all TRs). We end up with a vector of correlation values for each participant ${\bf c}_p^m$ associated with each encoding-model $m$. Each entry in this vector corresponds to a brain region. We can average the prediction performance across participants to obtain an average encoding-model prediction performance $\boldsymbol \mu^m$.

\paragraph{Encoding-model performance variability}
For each encoding-model $m$ and each brain-region $i$ we measure the across participant performance variability. We first calculate the standard deviation ${\sigma}^{i,m}$ of the predictive performance ${c}^{i,m}$ across all participants. We term this standard deviation the "non-standardized variability". To evaluate the amount of variability relative to the predictive performance, we calculate the coefficient of variation. The coefficient of variation for each encoding-model $m$ and brain region $i$ is defined as $v^{i,m}$:

$$ \nu^{i,m} = \sigma^{i,m} / |\mu^{i,m}| $$

Where $\sigma^{i,m}$ is the standard deviation of the encoding-model performance across all participants, $|\mu^{i,m}|$ is the absolute value of the mean of the encoding-model performance across all participants. In other words, we normalize the standard deviation by the absolute value of the average of the predictive performance across all participants. We also present the non-standardized variability (i.e., the standard deviation) in Supplementary Fig.~\hyperref[fig:sup_voxel_feature_specific_non_norm_var]{5}.

\paragraph{Testing for significance of encoding-model performance:}
For each participant, feature-space matrix, and brain region combination we test if the predictive performance $c_p^{i,m}$ is significantly higher than chance. We first create an empirical null distribution by resampling 10,000 permutations for each combination. We create each distribution by permuting contiguous chunks of 20 seconds (20 TRs for the movie task and resting state feature-space matrix, corresponding to 20 seconds, and 28 TRs for the motor task feature-space matrix corresponding to 20.160 seconds) of predicted fMRI data, similar to the permutation test used by Deniz et al.\textsuperscript{\citep{Deniz2019TheModality}} to account for autocorrelation in voxel responses. Then we calculate the predictive performance for each of the 10,000 randomized predictions. These 10,000 values correspond to our empirical distribution of chance performance. We compute the empirical \textit{p}-value of the un-permuted predictive performance $c_p^{i,m}$ using this empirical distribution, by computing the proportion of values at least as large as $c_p^{i,m}$. We use the Benjamini-Hochberg procedure to control the False Discovery Rate (FDR) at $\alpha = 0.05$\textsuperscript{\cite{Benjamini1995ControllingTesting}}. We consider a participant, feature-space, brain region combination to have a predictive performance significantly higher than chance if the one-sided permutation test \textit{p}-value is < $0.05$, FDR corrected. 

\subsubsection*{Encoding-models and behavior}
\paragraph{Correlation of encoding-model performance and behavior measures:} 
We calculate the correlation between each of the 15 cognitive measure scores and encoding-model performance. We use Pearson correlation. We use each of the three encoding-models for the video clip and one encoding-model for the motor task. For each ROI, the prediction performance across participants is correlated with their cognitive score to obtain the within-set correlation. 

\paragraph{Out-of-set prediction:}
We calculate if individual differences in predictive performance predict individual differences in cognitive behavior in an out-of-set setup, where the model is evaluated on held-out data. We estimate behavior models which predict a participant's cognitive measure score as a function of a vector of ROI or voxel predictive performance for a specific encoding-model. Because some areas of the brain might not be reliably predicted and thus will only constitute noisy features, we perform a simple feature selection step without considering the measures that we want to predict. For each encoding-model the vector of predictive performance contains all the ROIs/voxels where the performance was predicted significantly higher than chance (one-sided permutation test, alpha = $0.05$, FDR corrected) in at least a third of the participants (30 out of 90).  We use each of the three encoding-models for the video clip task and one encoding-model for the motor task. We use ridge regression to estimate these behavior models. For each cognitive measure the ridge regularization parameter is selected independently using nested 10-fold-cross-validation. 

For each cognitive measure $b$, we fit a model to predict the cognitive measure score $s_p^b$ of participant $p$ as a function of participant $p$'s prediction performance ${\bf c}_p^m$ for a given encoding-model $m$. For each encoding-model, $m$ is associated with a matrix ${\bf C}^m$ where each row is a vector of values ${\bf c}_p^m$ for participant $p$, one value for each brain region (i.e., ROI, voxel) of participant $p$. We estimate the model performance across participants by using a leave-one-family-out approach, similar to previous work\textsuperscript{\citep{Jiang2020Task-inducedRelationships, Dubois2018Resting-StateExperience, Feilong2021TheTopographies}}. The 90 participants were from 45 families, thus this cross-validation approach split the entire dataset into 45 folds. For each fold: 
\begin{enumerate}
    \item We split the predictive performance matrix ${\bf C}^m$ and cognitive measures scores ${\bf S}$ into training ${\bf C}^{R,m}$, ${\bf S}^R$ and validation matrices ${\bf C}^{V,m}$, ${\bf S}^V$. The test matrices contain data from $k$ individuals (from the same family), and the training matrices contain data from 90-$k$ individuals.
    \item We use the training matrices to estimate a model $U^{m}$:
    
   $$ \argminA_{U^{m}} || S^{R} - {\bf C}^{R,m}{\bf U}^{m} ||_{2}^{2} + \boldsymbol\lambda||{\bf U}^{m}||_{2}^{2} $$

    Similar to previously, we first identify the regularization parameter, $\lambda^{b}$ that minimizes the nested 10-fold cross-validation error for each cognitive measure $b$. The $\lambda^{b}$ are thus chosen independently for each score.  After the $\lambda^{b}$s are chosen, 
    we  estimate $\hat {\bf U}_p^{m}$ using all the training matrices.  
    \item We predict the validation cognitive scores as: $\hat{\bf S}_p^{V,m} = {\bf C}^{V,m}\hat{\bf U}^{m}$. 
\end{enumerate}

We then compute the prediction performance for each encoding-model predictive performance matrix, cognition score, by correlating the concatenation of the predicted scores across all folds with the observed scores across the same folds (i.e., Pearson correlation is performed across all 90 participants). We end up with correlation values $\eta_b^m$ for each score $b$ associated with each encoding-model $m$.

\paragraph{Chance performance of behavior model predictive performance:}
For each cognitive measure and encoding-model combination we create an empirical null distribution by resampling 10,000 permutations for each combination. We create each distribution by permuting predicted scores. 
Then we calculate the predictive performance for each of the 10,000 randomized predictions. These 10,000 values correspond to our empirical distribution of chance performance. For plotting purposes, we compute the 90\% confidence intervals for chance performance using the 5th and 95th quantiles of this empirical distribution. 
We also compute the empirical \textit{p}-value of the un-permuted predictive performance $\eta_b^m$ using this empirical distribution, by computing the proportion of values at least as large as $\eta_b^m$. We use the
Benjamini-Hochberg procedure to control the FDR at $\alpha = 0.05$\textsuperscript{\cite{Benjamini1995ControllingTesting}}. We consider a cognitive score, encoding model combination to have a predictive performance significantly higher than chance if the one-sided permutation test \textit{p}-value is < 0.05, FDR corrected.

\subsubsection*{ISC}
\label{ISC_methods}
 ISC measures how correlated a brain region is across different participants, most frequently calculated as the Pearson correlation of the brain activity between the same brain region for all pairs of participants\textsuperscript{\citep{Hasson2004IntersubjectVision, Nastase2019MeasuringCorrelation}}. We calculate ISC a measure of how correlated a brain region is across different participants. For each ROI we calculate the Pearson correlation of the brain activity between all pairs of participants. To obtain the ISC for each ROI, for each ROI we average the Fisher z-transformed Pearson correlation across all pairs of participants and then inverse Fisher-transform the average, similarly to previous works\textsuperscript{\citep{Hasson2010ReliabilityStimulation, Nastase2019MeasuringCorrelation, Jaaskelainen2016BrainHumor,  Gao2020ReliabilityImaging}}. These results are shown in Supplementary Fig.~\hyperref[fig:sup_ISC]{14}.

\subsubsection*{Definitions of FC and SC}
\label{FCSC_methods}
When we refer to FC and SC throughout the paper we are referencing the common definitions of these measures, which are as follows. FC measures the network interactions between brain regions, most frequently calculated as the Pearson correlation of the fMRI BOLD response between each pair of brain regions\textsuperscript{\citep{Friston1993FunctionalSets, Biswal1995FunctionalMrib, Mohanty2020RethinkingExtraction}}. SC measures the white matter connections between brain regions, and is typically calculated as diffusion MRI fiber pathways\textsuperscript{\citep{Wedeen2008DiffusionFibers}} or local connectome fingerprints\textsuperscript{\citep{Yeh2016Connectometry:Connectome}}. 

\subsubsection*{Identifying visual and language ROIs on the cortical surface}
We identify Shen atlas\textsuperscript{\citep{Finn2015FunctionalConnectivity, Shen2013GroupwiseIdentification}} ROIs involved with processing language and visual information. These ROIs are selected based on previously identified language and visual regions, which are entirely independent of our data. The visual cortex processes visual information. As Brodmann areas 17-19\textsuperscript{\citep{brodmann1909vergleichende}} comprise the visual cortex, the 25 Shen atlas ROIs with these labels were considered visual ROIs. The Brodmann area labels were provided on the BioImage Suite Web (\url{bioimagesuiteweb.github.io/webapp/connviewer.html})\textsuperscript{\citep{Shen2017UsingConnectivity}}. Regions of the brain involved with processing language-relevant information were previously identified by Fedorenko et al.\textsuperscript{\citep{Fedorenko2010NewSubjects}}. We obtained the left-hemisphere language ROIs from \url{https://evlab.mit.edu/funcloc/download-parcels} and mirrored these to obtain the right-hemisphere ROIs. We consider a Shen atlas ROI to be a language ROI if it is within a region that processes language-relevant information. This results in 27 Shen atlas ROIs that are language ROIs. 

\subsubsection*{Visualization on the cortical surface}
We use Pycortex\textsuperscript{\citep{Gao2015Pycortex:FMRI}} to visualize data on the brain. Pycortex creates the projection by sampling onto the cortical surface reconstruction. We project onto the MNI152 T1 2mm template in figures where the motor task results are presented to keep the data resolution consistent for comparison purposes, and MNI152 T1 1.6mm template otherwise.   

\newpage
\section*{Author contributions}
J.W. and L.W. conceptualized the research and created the statistical framework. J.W. performed analysis. J.W. and L.W. interpreted the data. J.W. wrote the first draft of the manuscript. J.W. and L.W. edited the manuscript together. 

\section*{Competing interests statement}
The authors declare no competing interests.

\section*{Acknowledgements}
The authors thank the Human Connectome Project project for making available the fMRI data and the behavioral measures, the Gallant lab for creating the visual features for the short movies, and Emily Finn for releasing code relevant to ROI preprocessing. The authors would like to also thank Tim Verstynen, Mariya Toneva, Michael Tarr, Jessica Wisnowski and Aaditya Ramdas for discussion of and comments on the manuscript. 

\section*{Funding Sources}
JW and LW were funded by startup funds from the Machine Learning Department at Carnegie Mellon University. JW was funded by the Center for Machine Learning and Health, and LW by the Google Faculty Research Award.

\newpage

\bibliography{references, moreRefs}
\bibliographystyle{unsrtnat} 
\addcontentsline{toc}{section}{Bibliography}

\newpage
\section*{Supplementary Material}
\setcounter{figure}{0} 
\captionsetup[figure]{labelfont=bf, labelformat={default}, name={Supplementary Figure},labelsep=period}
\begin{figure}[h]
    \centering
    \includegraphics[width=0.9\textwidth]{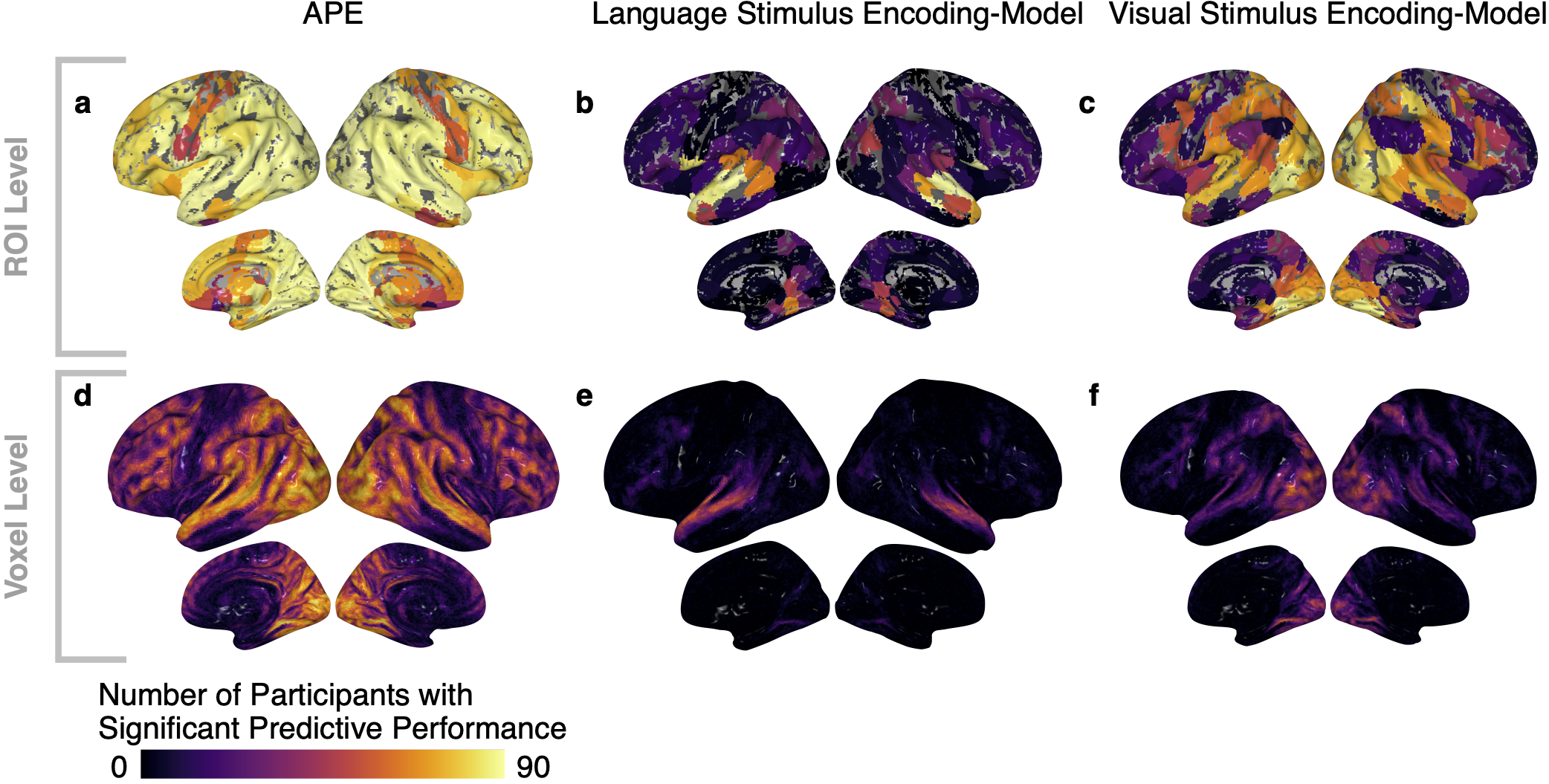}
    \begin{flushleft}
    \caption{\textbf{Video clip task encoding-models' predict BOLD response in brain regions significantly higher than chance.} \textbf{a}-\textbf{f}, Number of participants for which predictive performance is significantly higher than chance for the video clip APE (\textbf{a,d}), language stimulus encoding-model (\textbf{b,e}), visual stimulus encoding-model (\textbf{c,f}) at the ROI (\textbf{a-c}) and voxel level (\textbf{d-f}) (one-sided permutation test, alpha = $0.05$, FDR corrected). For each of the three types of encoding-models' (APE, language stimulus encoding-model, visual stimulus encoding-model) the voxel level results align with ROI level results, but also reveal a concentrated area where a high number of participants have predictive performance significantly higher than chance in the superior temporal gyrus (APE and language stimulus encoding-model), and early visual cortex (APE and visual stimulus encoding-model). Additionally, there is a magnitude difference for the same region between the ROI and voxel level. This is likely because an ROI's BOLD response is an average of the BOLD response across voxels', therefore the ROI response is less noisy and consequently easier to predict than the voxel response. Further, the voxel results show greater spatial specificity than the ROI results.}  
    \end{flushleft}
    \label{fig:sup_sig_voxel_EM}
\end{figure}

\begin{figure}[h]
    \centering
    \includegraphics[width=1\textwidth]{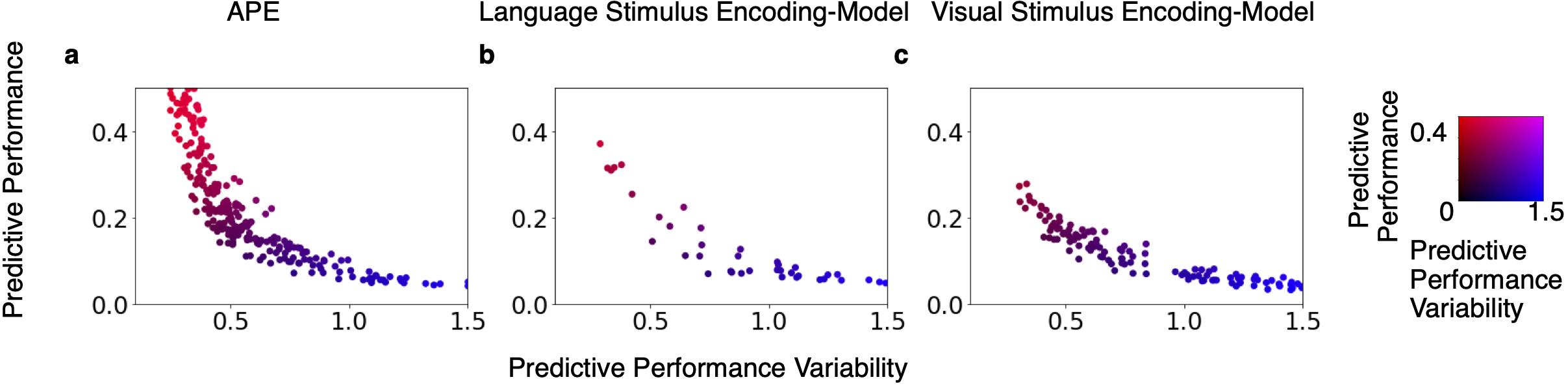}
    \begin{flushleft}
    \caption{\textbf{The variability of voxel level encoding-model performance is feature-space specific at an ROI level.} 
    \small \textbf{a}-\textbf{c}, In the scatter plots predictive performance variability versus predictive performance for the three video clip task encoding-models averaged across all participants are plotted per ROI. ROIs are colored according to a 2D color map, where red corresponds to high predictive performance, blue to high variability, purple to high predictive performance and variability, and black to low predictive performance and variability. Similarly to Fig.~\ref{fig:StimFeatSpecific}, our results show a strong trend for ROIs with high predictive performance to have low variability (red), and ROIs with low predictive performance to have high variability (blue). These ROI level results are consistent with the voxel level results in Fig.~\ref{fig:StimFeatSpecific}, and highlight that predictive performance depends on the type of encoding-model.} 
    \end{flushleft}
    \label{fig:sup_roi_satter}
\end{figure}

\begin{figure}[h]
    \centering
    \includegraphics[width=0.75\textwidth]{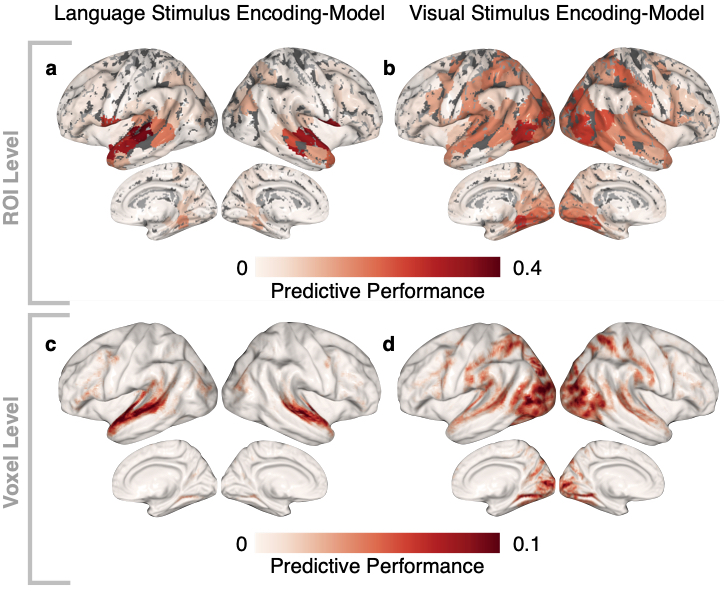}
    \begin{flushleft}
    \caption{\textbf{Stimulus encoding-model performance is feature-space specific.}
    \textbf{a-d}, Predictive performance of the video clip task language stimulus encoding-model (\textbf{a,c}), and visual stimulus encoding-model (\textbf{b,d}) averaged across all participants at an ROI (\textbf{a,b}) and voxel level (\textbf{c,d}). We find high predictive performance at both levels for the following: (\textbf{a,c}) language stimulus encoding-model in previously defined language regions\textsuperscript{\citep{Fedorenko2010NewSubjects}} including the anterior (ATL) and posterior temporal lobes (PTL); (\textbf{b,d}) visual stimulus encoding-model in the visual cortex, and the dorsal ("where") pathway and the ventral ("what") pathway, consistent with previous results\textsuperscript{\citep{Huth2012ABrainc}}.}
    \end{flushleft}
    \label{fig:sup_pred_perf_stimlus_models}
\end{figure}
\begin{figure}[h]
    \centering
    \includegraphics[width=0.6\textwidth]{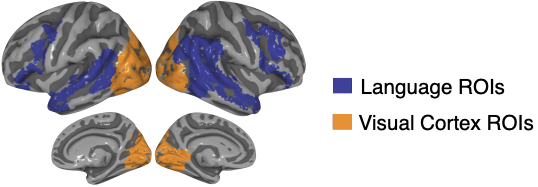}
    \begin{flushleft}
    \caption{\textbf{Language and visual cortex ROIs.} The ROIs are colored according to the legend. Blue corresponds to ROIs that overlap with previously defined language ROIs\textsuperscript{\citep{Fedorenko2010NewSubjects}}. Orange corresponds to ROIs that are in the visual cortex, specifically in Brodmann areas 17-19\textsuperscript{\citep{Shen2017UsingConnectivity}}.} 
    \end{flushleft}
    \label{fig:sup_roi_map}
\end{figure}

\begin{figure}[h]
    \centering
    \includegraphics[width=1\textwidth]{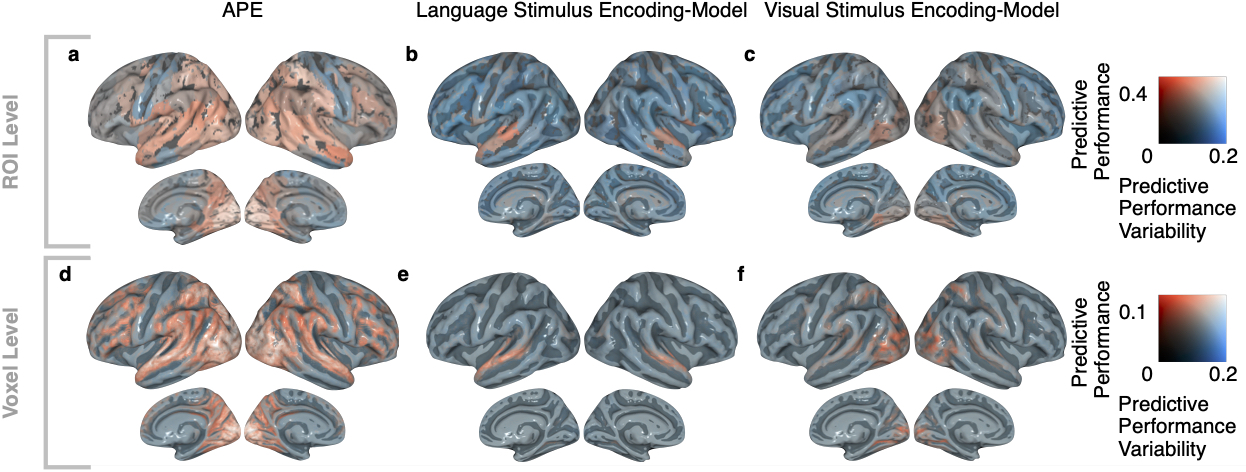}
    \begin{flushleft}
    \caption{\textbf{The non-standardized variability (standard deviation) of encoding-model performance is feature-space specific.} 
    \small \small \textbf{a}-\textbf{c}, Comparison of the average predictive performance to its standard deviation for the three video clip encoding-models at an ROI (\textbf{a-c}) and voxel level (\textbf{d-f}), formatted similarly to Fig.~\ref{fig:StimFeatSpecific}a-f. Similarly to Fig.~\ref{fig:StimFeatSpecific}, we find high predictive performance and low standard deviation at both levels for the following: (\textbf{a,d}) APE in previously defined language regions\textsuperscript{\citep{Fedorenko2010NewSubjects}} including the anterior (ATL) and posterior temporal lobes (PTL), the visual cortex, and the dorsal ("where") pathway and the ventral ("what") pathway; (\textbf{b,e}) language stimulus encoding-model in the same previously defined language regions; (\textbf{c,f}) visual stimulus encoding-model in the visual cortex, and the dorsal and ventral pathways, consistent with previous results\textsuperscript{\citep{Huth2012ABrainc}}. This suggests that the pattern of variability is consistent whether the variability is standardized as in Fig.~\ref{fig:StimFeatSpecific} or not.} 
    \end{flushleft}
    \label{fig:sup_voxel_feature_specific_non_norm_var}
\end{figure}

\begin{figure}[h]
    \centering
    \includegraphics[width=0.8\textwidth]{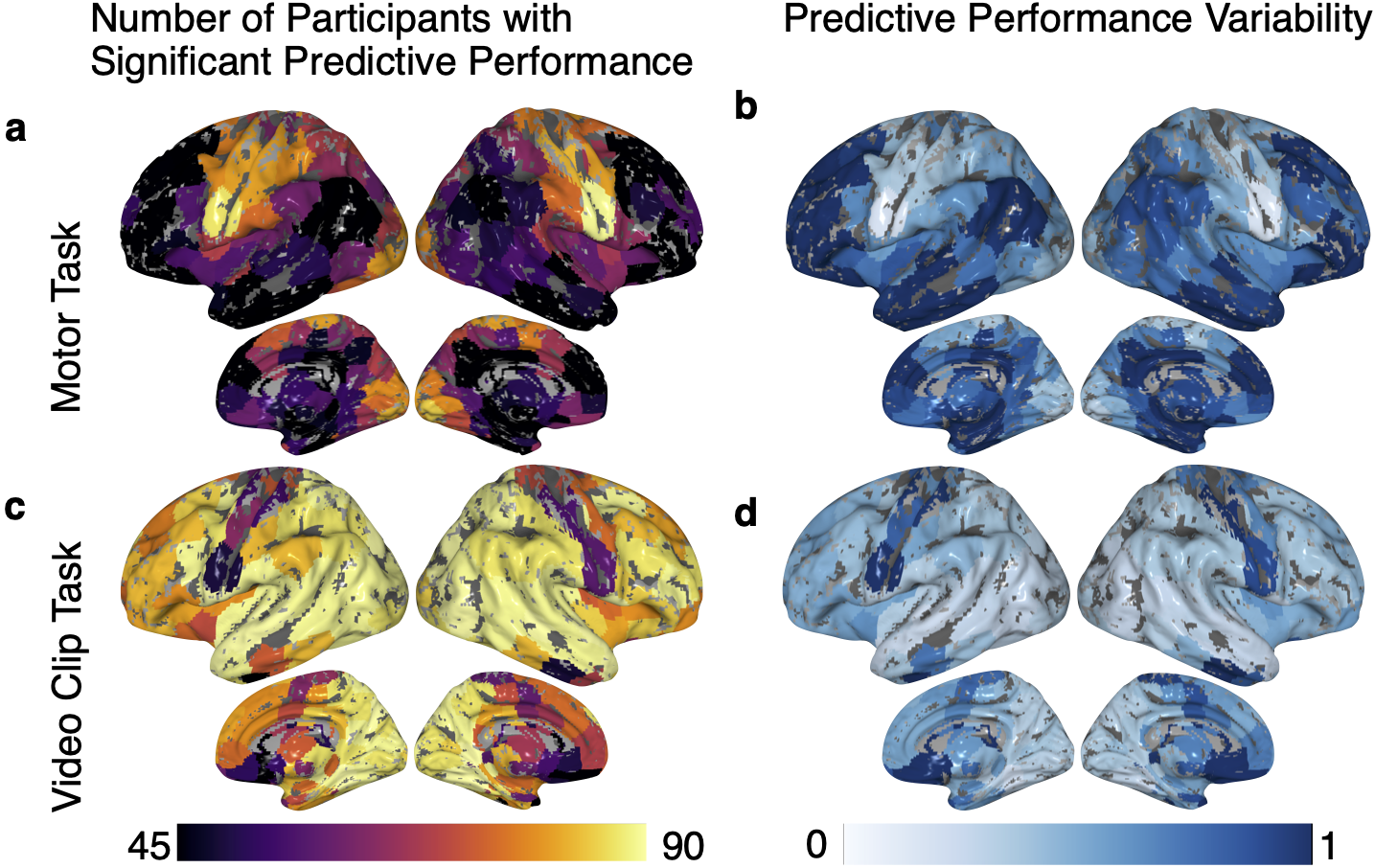}
    \begin{flushleft}
    \caption{\textbf{The variability of ROI level encoding-model performance is task-specific.} \small Comparison of the motor task and video clip task at an ROI level, formatted similarly to Fig.~\ref{fig:StimSpecific}e-h. \textbf{a,c}, Number of participants for which the predictive performance is significantly higher than chance for the motor task (\textbf{a}) and video clip task (\textbf{c}) (one-sided permutation test, alpha = $0.05$, FDR corrected) is task-specific. \textbf{b,d} Predictive performance variability during the motor task (\textbf{b}) is low in the motor cortex and high elsewhere, but during the video clip task (\textbf{d}), it is low in the visual, temporal, and prefrontal cortices and high elsewhere. Predictive performance magnitude and variability are task-specific. Further, these ROI level results are consistent with the voxel level results in Fig.~\ref{fig:StimSpecific}.} 
    \end{flushleft}
    \label{fig:sup_roi_motor}
\end{figure}

\begin{figure}[h]
    \centering
    \includegraphics[width=.4\textwidth]{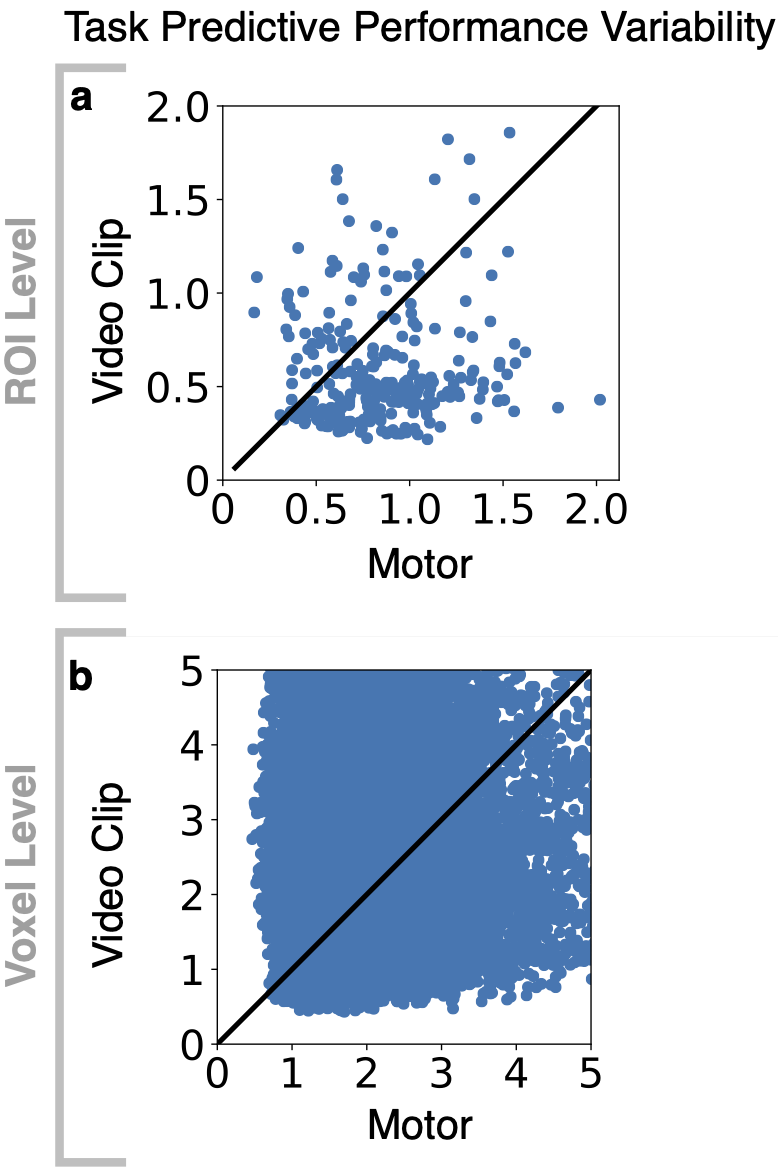}
    \begin{flushleft}
    \caption{\textbf{Video clip task and motor task APE performance variability is not correlated.} In the scatter plots predictive performance variability of the motor task APE versus the video clip task APE are plotted per ROI or voxel. The black diagonal line denotes $y$=$x$, if points are above this line those ROIs/voxels video clip task performance variability is greater than their motor task variability. We find that the variability between the two tasks is not correlated (ROI level: correlation = $0.04$, \textit{p} = $0.54$, voxel level: correlation = $-0.0004$, \textit{p} = $0.93$).}     
    \end{flushleft}
    \label{fig:supp_motor_movie_var}
\end{figure}

\begin{figure}[h]
    \centering
    \includegraphics[width=1\textwidth]{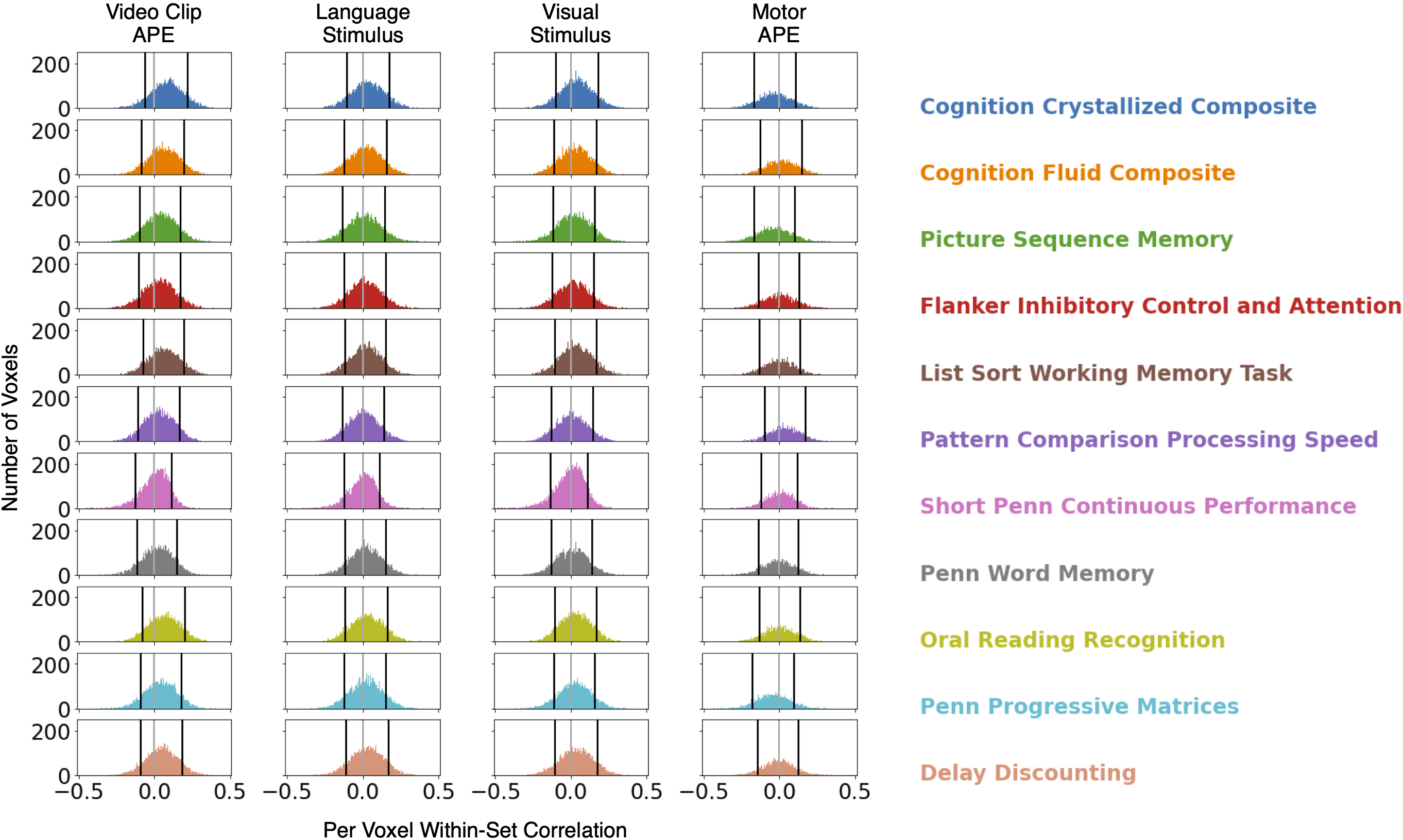}
     \begin{flushleft}
     \caption{\textbf{Individual differences in encoding-model performance are associated with cognitive measures at a voxel level.} Within-set correlation of encoding-model performance with scores from 11 cognitive measures, formatted similarly to Fig.~\ref{fig:behavior}. Similarly to Fig.~\ref{fig:behavior}, we observe variability in the distributions of correlations across different encoding-model/cognitive measure combinations. For some encoding-model/cognitive measure combinations, there is a clear positive skew of the correlations: for example, there is a positive correlation in most of the brain between a participant's video clip APE predictive performance and their \emph{cognition crystallized composite} score. However, there does not seem to be a consistent pattern between the same encoding-model's performance, and the participants' \emph{short penn continuous performance} score.}  
     \end{flushleft}
    \label{fig:sup_corr_cogn_voxel}
\end{figure}

\begin{figure}[h!]
    \centering
    
    \includegraphics[width=1\textwidth]{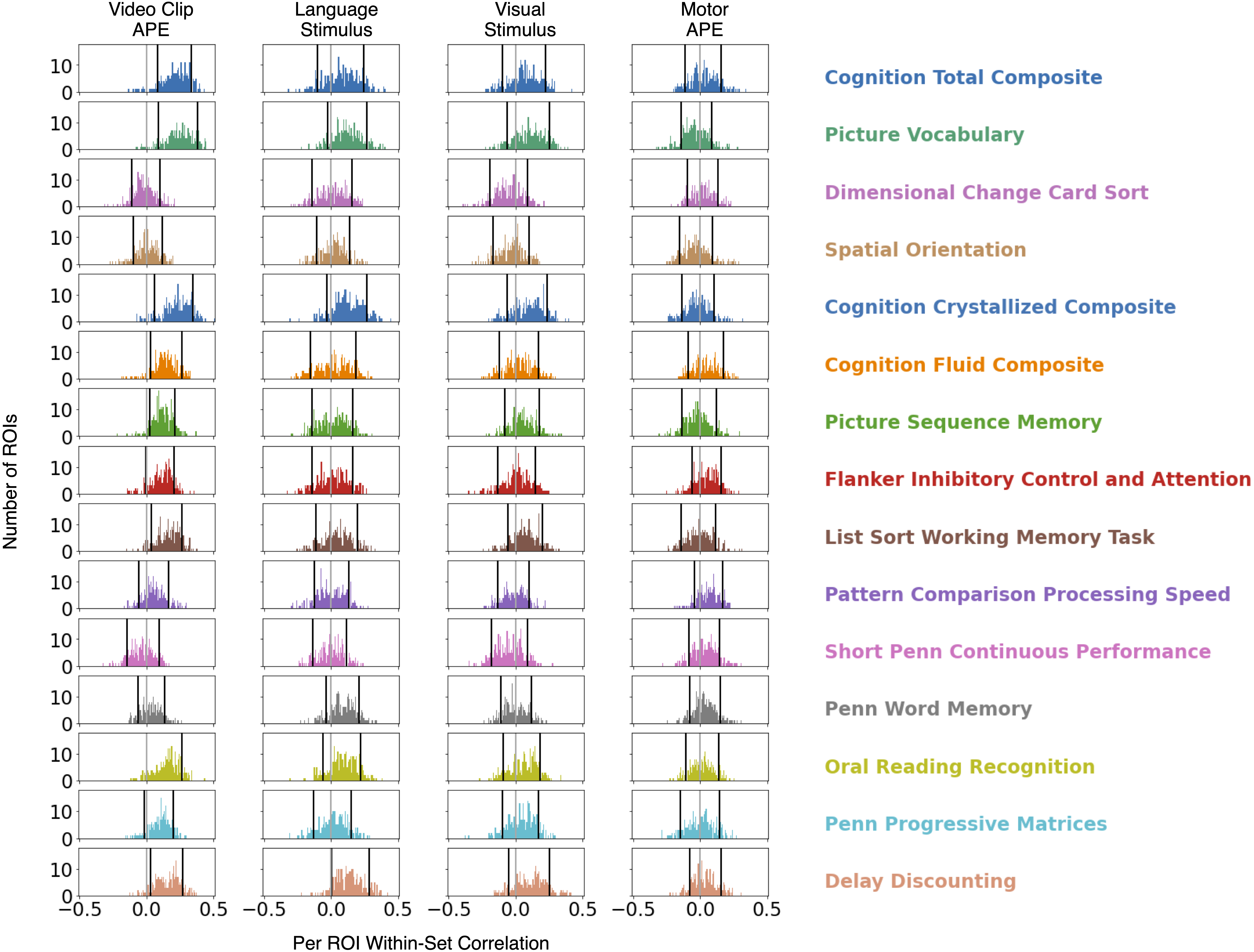}
    \begin{flushleft}
    \caption{\textbf{Cognitive behavior is associated with individual differences in encoding-model performance at an ROI level.}}
    \small Correlation of encoding-model performance at an ROI level with scores from 15 cognitive measures, formatted similarly to Fig.~\ref{fig:behavior}. We use all four encoding-models (three for the video clip task and one for the motor task). Similarly to Fig.~\ref{fig:behavior}, we observe variability in the distributions of correlations across different encoding-model/cognitive measure combinations. For some encoding-model/cognitive measure combinations, there is a clear positive skew of the correlations. For example, there is a positive correlation in most of the brain between a participant's video clip APE predictive performance and their \emph{cognition total composite} score. However, the pattern between the same encoding-model's performance and the participants' \emph{spatial orientation} scores is inconsistent. Further, these ROI level results are consistent with the voxel level results in Fig.~\ref{fig:behavior} and Supplementary Fig.~\hyperref[fig:sup_corr_cogn_voxel]{8}.
    \end{flushleft}
    \label{fig:sup_corr_cogn_roi}
\end{figure}

\begin{figure}[h]
    \centering
    \includegraphics[width=1\textwidth]{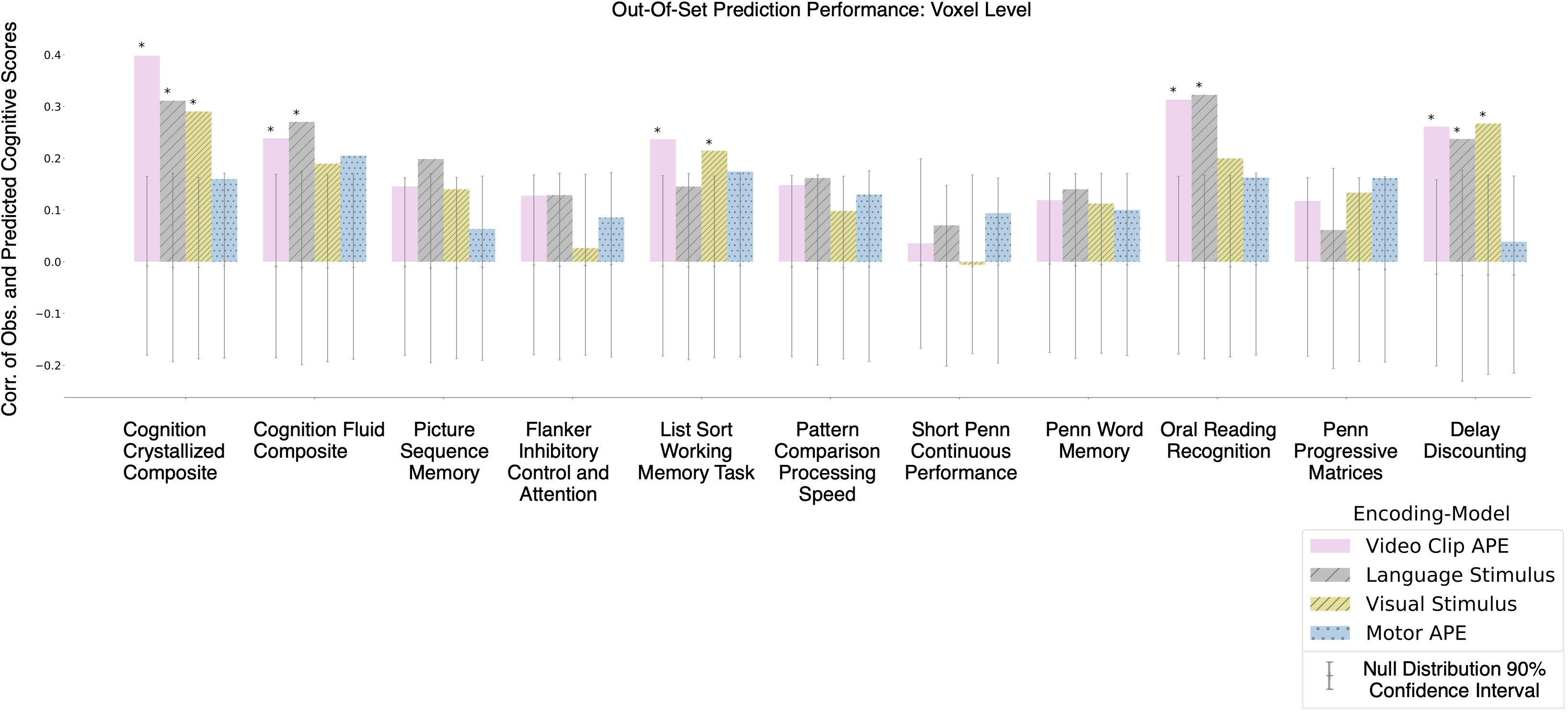}
     \begin{flushleft}
     \caption{\textbf{Individual differences in encoding-model performance at a voxel level predict cognitive measures.} Out-of-set predictive performance of participant cognitive scores, formatted similarly to Fig.~\ref{fig:behavior_pred}. We find that all three-video clip task encoding-models’ performances predict two measures of cognition (\emph{cognition crystallized composite}, \emph{delay discounting}) significantly higher than chance. We also find that the three-video clip task encoding-models' performance predict the same cognitive measures differently: for example, only the video clip APE and language stimulus encoding-model predict \emph{cognition fluid composite} and \emph{oral reading recognition} significantly higher than chance. However, both the video clip APE and visual encoding-model predict the \emph{list sort working memory task} significantly higher than chance. The motor task encoding-models' performance does not predict any of the 11 measures significantly higher than chance. These task and encoding-model specific predictive capabilities reveal that individual differences in processing different tasks and encoding-models can predict different behavioral measures. The dark gray error bars denote a 90\% confidence interval for chance performance (from the 5th to the 95th quantile). * \textit{p} < 0.05 (one-sided permutation test, alpha = $0.05$, FDR corrected)} 
     \end{flushleft}
    \label{fig:behav_predict_voxel_11}
\end{figure}

\begin{figure}[h!]
    \centering
    \includegraphics[width=0.8\textwidth]{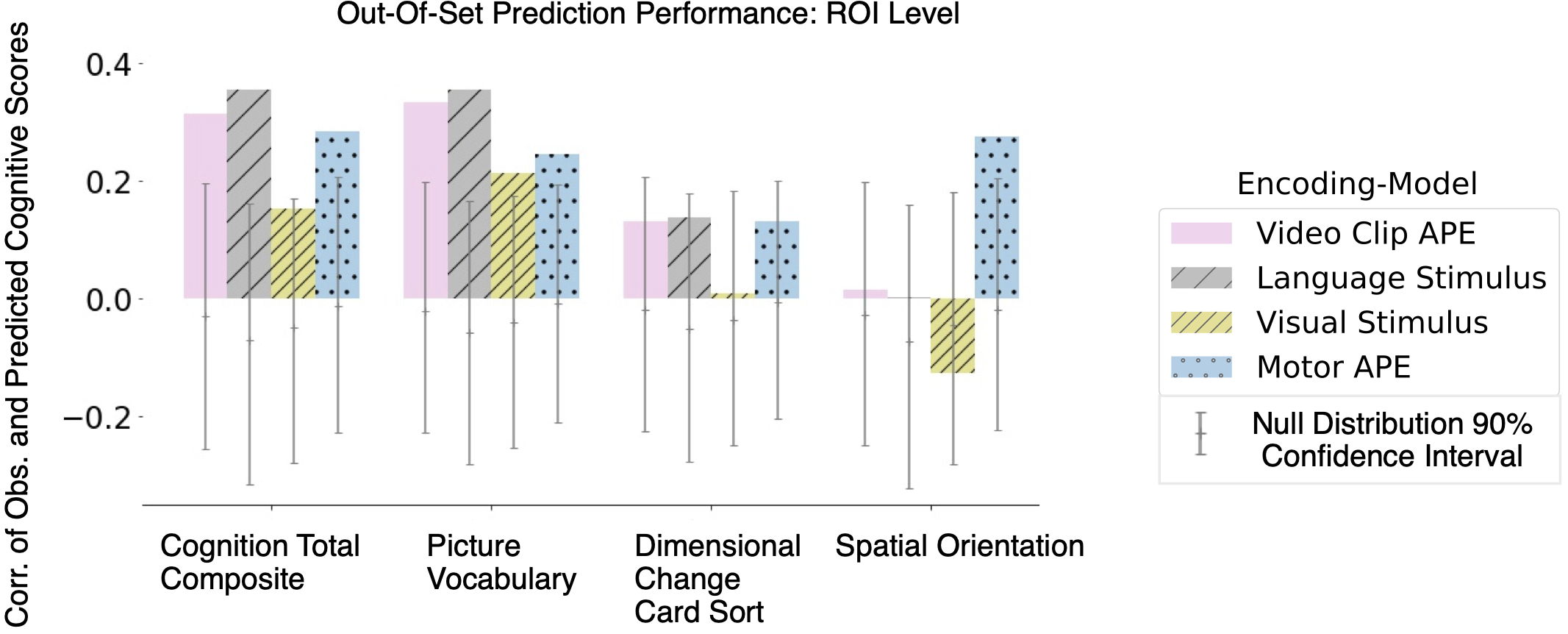}
    \begin{flushleft}\caption{\textbf{Individual differences in encoding-model performance at an ROI level predict cognitive behavior.}}
    \small Out-of-set predictive performance of participant cognitive scores, formatted similarly to Fig.~\ref{fig:behavior_pred}. Performance was measured as the correlation between actual and predicted participant cognitive measure scores. Each participant's score was predicted from a vector of ROI encoding-model performance (using all four encoding-models). Predictive performance was evaluated on held-out participants in a leave-one-family-out-cross-validation setup.  We find that all four encoding-models' performances predict the \emph{picture vocabulary} score higher than the 95\% quantile of an empirical distribution estimating chance performance, estimated through a permutation test. The motor task encoding-models’ performances predict the \emph{spatial orientation} measure (variable short penn line orientation) higher than the 95\% quantile of chance performance. These task-specific predictive capabilities suggest that individual differences in processing different tasks can predict different behavior measures. Further, task-specific predictive capabilities (at an ROI level) are similar to the voxel level results in Fig.~\ref{fig:behavior_pred}, however the magnitude was not found to be significantly higher than chance after multiple comparison correction, likely due to a loss of power when averaging the voxel data into ROIs. The dark gray error bars denote a 90\% confidence interval for chance performance (from the 5th to the 95th quantile). 
    \end{flushleft}
    \label{fig:behavior_pred_roi_4_measures}
\end{figure}

\begin{figure}[h]
    \centering
    \includegraphics[width=1\textwidth]{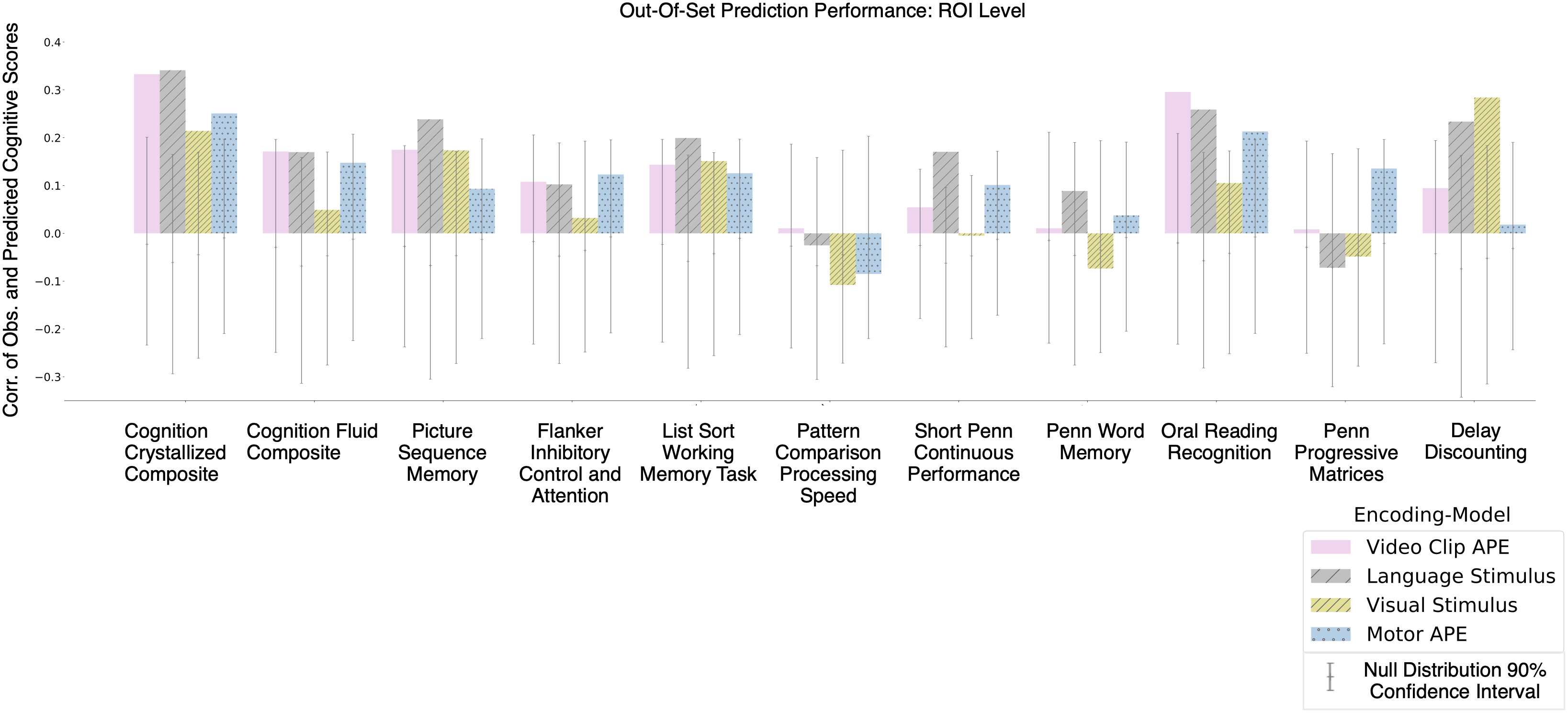}
    \begin{flushleft}
     \caption{\textbf{Individual differences in encoding-model performance at an ROI level predict cognitive measures.} Out-of-set predictive performance of participant cognitive scores, formatted similarly to Supplementary  Fig.~\hyperref[fig:behav_predict_voxel_11]{10}. We find that all four encoding-models predict the \emph{cognition crystallized composite} measure higher than the 95\% quantile of an empirical distribution estimating chance performance, estimated through a permutation test. We also find that the three-video clip task encoding-models' performance predict the same cognitive measures differently: for example, only the language stimulus encoding-model predicts the \emph{list sort working memory} measure higher than the 95\% quantile of chance performance. However, both the video clip APE and language stimulus encoding-model predict the \emph{oral reading recognition} measure higher than the 95\% quantile of chance performance. These encoding-model specific predictive capabilities reveal that individual differences in processing different feature-spaces of the same task can predict different behavioral measures. Further, these encoding-model specific predictive capabilities (at an ROI level) are similar to the voxel results in Supplementary Fig.~\hyperref[fig:behav_predict_voxel_11]{10}, however the magnitude was not found to be significantly higher than chance after multiple comparison correction, likely due to a loss of power when averaging the voxel data into ROIs. The dark gray error bars denote a 90\% confidence interval for chance performance (from the 5th to the 95th quantile).} 
    \end{flushleft}
    \label{fig:behav_predict_roi_11}
\end{figure}

\begin{figure}[h]
    \centering
    \includegraphics[width=0.4\textwidth]{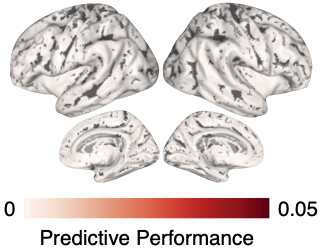}
    \begin{flushleft}
    \caption{\textbf{Resting state encoding-model performance is negligible.} Predictive performance of the resting state task APE averaged across all participants at an ROI level. Prediction performance is negligible across the entire brain. This demonstrates the inability to study information processing from resting-state data. Consequently, as expected, our framework is not applicable to resting-state data.}
    \end{flushleft}
    \label{fig:sup_rest}
\end{figure}

\newpage
\begin{figure}[h]
    \centering
    \includegraphics[width=0.75\textwidth]{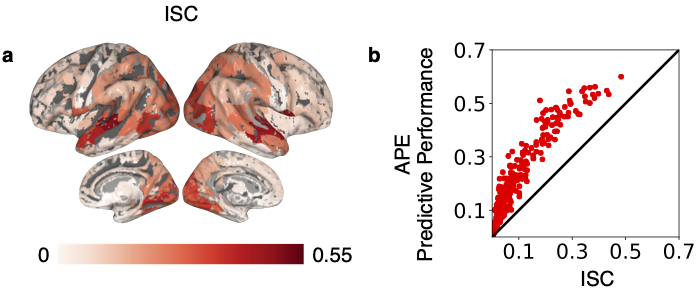}
    \begin{flushleft}
    \caption{\textbf{Video clip ISCs are not as strong as out-of-set predictive performances.} \textbf{a}, ISC of the video clip task at a ROI level. ISC is measured as the Pearson correlation of the brain activity between the same brain region for all pairs of participants. ISC is high in ROIs in the language ROIs (ATL, PTL) and visual cortex. \textbf{b}, In the scatter plot the video clip ISC versus the video clip APE predictive performance are plotted per ROI. The black diagonal line denotes $y$=$x$. Almost every point (267/268) is above this line, indicating that for those ROIs APE predictive performance is greater than their ISC. APE predictive performance is much better than ISC especially in the frontal cortex, which is understood to be more variable across participants than sensory areas (primary auditory, primary somatosensory, and primary visual cortices).} 
    \end{flushleft}
    \label{fig:sup_ISC}
\end{figure}

\FloatBarrier 

\newpage
\begin{table}[h]
    \centering
    \begin{flushleft}
    \caption{\textbf{HCP cognition measures, obtained from HCP S1200 data dictionary.} * Indicates measure from NIH Toolbox}
    \end{flushleft}
    \includegraphics[width=1\textwidth]{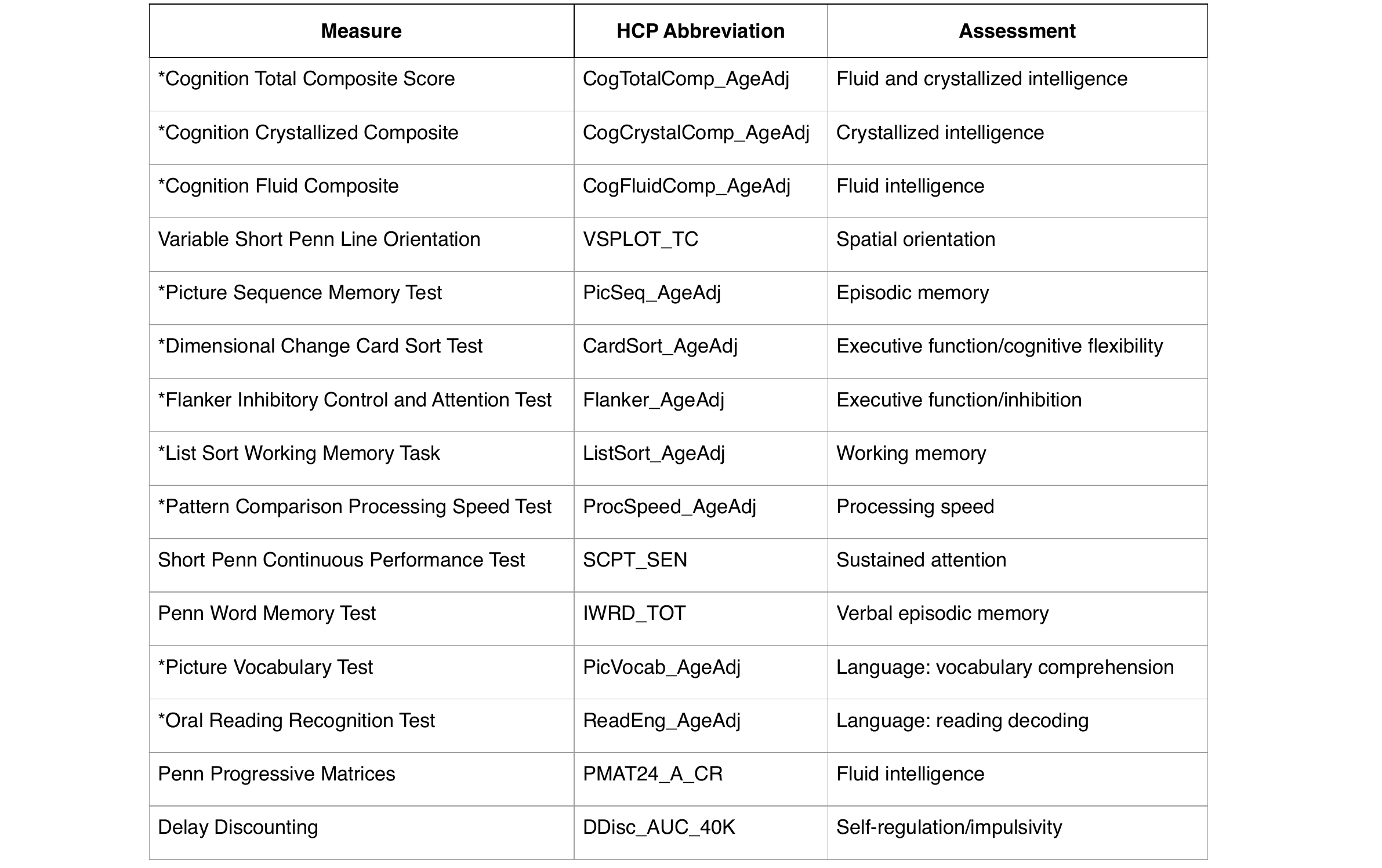}
    \label{tab:sup_cogn}
\end{table}

\end{document}